\documentclass[twocolumn,showpacs,preprintnumbers,amsmath,amssymb,APSl,prd,nofootinbib,superscriptaddress]{revtex4-1}

\usepackage{dcolumn}
\usepackage{bm}
\usepackage{ifpdf}
\usepackage{hyperref}
\usepackage{bm}
\usepackage{xcolor,color,graphicx,graphics}
\usepackage[spanish,english]{babel}
\usepackage[latin1]{inputenc}
\usepackage[OT1]{fontenc}
\usepackage{latexsym,amssymb,amsmath,amsfonts}
\usepackage{makeidx}
\usepackage{epsfig,subfigure}
\usepackage{natbib}
\usepackage{epstopdf}
\usepackage{mathrsfs}
\usepackage{hyperref}
\hypersetup{colorlinks=true, linkcolor=blue, citecolor=green}
\usepackage{enumerate}

\definecolor{red}{rgb}{1,0,0}

\def\+{^\dagger}

\def\<{\leftarrow}
\def\>{\rightarrow}

\def\({\left(}
\def\){\right)}

\newcommand{\LL}{{\cal L}}

\newcommand{\bi}{\begin{itemize}} 				\newcommand{\ei}{\end{itemize}}
\newcommand{\benu}{\begin{enumerate}} 		\newcommand{\enu}{\end{enumerate}}
\newcommand{\bd}{\begin{dinglist}{0}}     \newcommand{\ed}{\end{dinglist}}
\newcommand{\bfig}{\begin{figure}[htbp]}  \newcommand{\efig}{\end{figure}}
        			
\newcommand{\bc}{\begin{center}} 				  \newcommand{\ec}{\end{center}}
\newcommand{\be}{\begin{equation}} 				\newcommand{\ee}{\end{equation}}
\newcommand{\bsub}{\begin{subequations}}  \newcommand{\esub}{\end{subequations}}
\newcommand{\ben}{\begin{eqnarray}} 			\newcommand{\een}{\end{eqnarray}}
\newcommand{\ba}[1]{\begin{array}{#1}} 		\newcommand{\ea}{\end{array}}
\newcommand{\bea}{\begin{equation}\begin{array}{rcl}}
\newcommand{\eea}{\end{array}\end{equation}}

\begin{document}
\title{Rotating black holes in Eddington-inspired Born-Infeld gravity: an exact solution}

\author{Merce Guerrero} \email{merguerr@ucm.es}
\affiliation{Departamento de F\'isica Te\'orica and IPARCOS,
	Universidad Complutense de Madrid, E-28040 Madrid, Spain}
	
\author{Gerardo Mora-P\'{e}rez} \email{moge@alumni.uv.es}
\affiliation{Departamento de F\'{i}sica Te\'{o}rica and IFIC, Centro Mixto Universidad de Valencia - CSIC.
	Universidad de Valencia, Burjassot-46100, Valencia, Spain}

\author{Gonzalo J. Olmo} \email{gonzalo.olmo@uv.es}
\affiliation{Departamento de F\'{i}sica Te\'{o}rica and IFIC, Centro Mixto Universidad de Valencia - CSIC.
	Universidad de Valencia, Burjassot-46100, Valencia, Spain}
\affiliation{Departamento de F\'isica, Universidade Federal da
	Para\'\i ba, 58051-900 Jo\~ao Pessoa, Para\'\i ba, Brazil}
	
	\author{Emanuele Orazi} \email{orazi.emanuele@gmail.com}
\affiliation{ International Institute of Physics, Federal University of Rio Grande do Norte,
Campus Universit\'ario-Lagoa Nova, Natal-RN 59078-970, Brazil}
\affiliation{Escola de Ciencia e Tecnologia, Universidade Federal do Rio Grande do Norte, Caixa Postal 1524, Natal-RN 59078-970, Brazil}

\author{Diego Rubiera-Garcia} \email{drubiera@ucm.es}
\affiliation{Departamento de F\'isica Te\'orica and IPARCOS,
	Universidad Complutense de Madrid, E-28040 Madrid, Spain}

\date{\today}
\begin{abstract}
We find an exact, rotating charged black hole solution within Eddington-inspired Born-Infeld gravity. To this end we employ a recently developed correspondence or {\it mapping} between modified gravity models built as scalars out of contractions of the metric with the Ricci tensor, and formulated in metric-affine spaces (Ricci-Based Gravity theories) and General Relativity. This way, starting from the Kerr-Newman solution, we show that this mapping bring us the axisymmetric solutions of  Eddington-inspired Born-Infeld gravity coupled to a certain model of non-linear electrodynamics. We discuss the most relevant physical features of the solutions obtained this way, both in the spherically symmetric limit and in the fully rotating regime. Moreover, we further elaborate on the potential impact of this important technical progress for bringing closer the predictions of modified gravity with the astrophysical observations of compact objects and gravitational wave astronomy.
\end{abstract}

\maketitle

\section{Introduction} \label{sec:II}

Among the many known exact solutions of the field equations of Einstein's General Relativity (GR), only a handful  are known to carry actual physical meaning \cite{ExactBook}. Prominent among them is the family of stationary axisymmetric solutions. Indeed, the unicity's theorems \cite{UT1,UT2,UT3} tell us that the most general electrovacuum such solution is the Kerr-Newman one, which has been consistently interpreted as the external gravitational field to a body described solely by mass, electric charge and angular momentum \cite{Kerr:1963ud,Newman:1965my}. The reliability of the Kerr (\textit{i.e.} uncharged) solution to describe astrophysical black holes has been established by several means, including X-ray spectroscopy from the accretion disk of both stellar and supermassive black holes \cite{Jiang:2014loa,Cardenas-Avendano:2016zml} as well as via strong gravitational lensing and shadows (for a recent review see e.g. \cite{Cunha:2018acu}), recently culminated in the imaging of the shadow of the supermassive central object of the M87 galaxy \cite{M87}. Furthermore, the gravitational wave output resulting from the numerical analysis of the merger of two black holes described by the Kerr solution is compatible with the waveform signals observed by the LIGO-VIRGO Collaboration \cite{Abbott:2016blz,Abbott:2017oio}.

With the awakening of gravitational wave astronomy the possibility of testing the strong  field regime of the gravitational field and exploring for hints of new Physics beyond that of GR is closer than ever before \cite{Abbott:2018lct}. The correct identification and interpretation of such new physics regarding generation and propagation of gravitational waves in black hole mergers \cite{Yunes:2016jcc,Berti:2018vdi,Berti:2015itd}, for instance, requires a deep understanding of the underlying phenomena and, for this reason, obtaining exact analytical rotating solutions in different gravity theories and astrophysical scenarios is a pressing topic in the field. However, progress in this direction has met with some difficulties. Firstly, given the additional complexities typically present within the field equations of most modified theories of gravity (\emph{e.g.}, their fourth-order nature and/or their highly non-linear character), solving them  has proven a daunting challenge, typically requiring the introduction of additional simplifications such as constant curvature solutions, slowly-rotating scenarios or via numerical approximations, see \emph{e.g.} \cite{Konno:2009kg,Yunes:2009hc,Pani:2009wy,Pani:2011gy,Kleihaus:2011tg,Bambi:2013ufa,Cembranos:2011sr,Moffat:2014aja,Ayzenberg:2014aka,Maselli:2015tta,Bueno:2017hyj,Buoninfante:2018xif,Anabalon:2018qfv,Cano:2019ore,Ding:2019mal,Jusufi:2019caq} for some works on the subject\footnote{An alternative approach are model-independent parametrizations of the deviations from the Kerr black hole, which has some advantages as well as some limitations, see \emph{e.g.} \cite{Johannsen:2011dh,Konoplya:2016jvv}.}. Secondly, the development of advanced numerical codes to address this issue is tightly attached to the structure of the GR field equations, thus making the task of adapting them to the structure of the field equations of each proposal for modified gravity on a case-by-case basis extremely expensive from the point of view of computational and human resources. This state-of-the-art troubles the extraction of useful information from different modified theories of gravity in order to discriminate their predictions from those of GR, for instance, via waveforms/quasi-normal mode spectrum in current/future observational facilities such as LISA \cite{Berti:2005ys,Barausse:2020rsu}, or via X-ray spectroscopy in accretion disks around stellar-mass black holes \cite{Bambi:2016sac,Cardenas-Avendano:2019pec}.

The main aim of this paper is to put to work a novel technical improvement to find an exact rotating black hole solution of modified gravity. The theories of gravity under consideration are defined via Lagrangian densities built out of contractions of the Ricci tensor and the metric tensor (Ricci-Based Gravity theories, or RBGs for short) and formulated in metric-affine spaces, where metric and affine connection are independent entities \cite{Olmo:2011uz}. Unlike the usual metric formulation,  the metric-affine framework guarantees for these theories the second-order and ghost-free character of the corresponding field equations, which propagate only the two tensorial modes of the gravitational field at the speed of light, being thus automatically compatible with the reported data resulting from the merger of binary neutron stars systems \cite{AbbottNS}. 

The rotating black holes presented here are obtained using a recently established correspondence or \emph{mapping} between RBGs and GR, by which the field equations and the solutions of the later coupled to some matter Lagrangian can be mapped into the field equations and solutions of the former coupled to the same matter fields but with a different  Lagrangian density and vice versa \cite{Afonso:2018bpv}.  The power of this method is apparent. Instead of solving the highly non-linear field equations of the original RBG, we can transfer the problem to the GR side, use the whole known toolkit of analytical and numerical methods developed there to find a solution, and then map it back to the original RBG via purely algebraic transformations. The \emph{proof of concept} of this correspondence has been explicitly established for electromagnetic \cite{Afonso:2018mxn,Delhom:2019zrb} and scalar \cite{AOOR18c} fields, where the form of some known exact solutions was obtained, and more recently employed to find new compact objects in RBGs starting from a seed canonical scalar field solution of GR \cite{Afonso:2019fzv}.

In this work we shall use the Kerr-Newman solution of GR as the seed to obtain the corresponding solution in the RBG side, for which we choose the Eddington-inspired Born-Infeld (EiBI) theory of gravity \cite{BI0}. This pick is motivated due to the many applications of this model in astrophysics and cosmology \cite{BI1,BI2,BI3,BI4,BI5,BI6,BI7,BI8,BI9,BI10,BI11,Boehmer:2020hkn} (see \cite{BeltranJimenez:2017doy} for a review). We shall find that the matter source of GR, namely, Maxwell electromagnetism, is mapped on the EiBI side into a Born-Infeld-type electrodynamics \cite{BI34}. Though from a physical viewpoint this scenario is perhaps not the most interesting one (this would correspond to finding the matter counterpart on the GR side of EiBI gravity coupled to Maxwell electrodynamics), it will serve, on the one hand, to check the reliability of this method when applied to axisymmetric scenarios and, on the other hand, to begin exploring the qualitative differences of these solutions of modified gravity as compared to those of the Kerr-Newman black hole. Let us point out that rotating black holes are expected to discharge due to several effects \cite{Gibbons:1975kk,WaldBook} and, therefore, from a purely astrophysical perspective one can safely neglect the residual charge as a good approximation \cite{Bambi:2019xzp}. However, there are several mechanisms by which a residual (if tiny) charge would play a relevant physical role, for instance, on the interaction of the black hole with their accretion disks  \cite{Trova:2018bsf} or with cosmic rays passing in their vicinity \cite{Zajacek:2019kla}. Within RBGs, the role of the charge is to feed the new dynamics engendered by the new couplings with the matter fields ascribed to these theories. We shall see that, despite these limitations, the resulting solution will give us valuable information on the new physics introduced by these modifications of GR.

This work is organized as follows: in Sec. \ref{sec:II} we introduce the family of RBGs, construct the map with GR for anisotropic fluids, and particularize it for the case of EiBI gravity. In Sec. \ref{sec:III} electrostatic, spherically symmetric solutions are constructed via direct resolution of the field equations and by application of the mapping, verifying their mutual consistence and finding that GR coupled to Maxwell electrodynamics maps into EiBI gravity coupled to Born-Infeld electrodynamics. Subsequently, in Sec. \ref{sec:IV} we apply the same map to axisymmetric solutions finding an exact analytical rotating black hole solution, and characterize its most prominent features. Sec. \ref{sec:V} contains our conclusions and some discussion.

\section{The mapping method} \label{sec:II}

\subsection{Ricci-based gravities}

Let us begin by defining the family of gravitational theories we are interested in. These are given by an action of the form
\begin{eqnarray} \label{eq:actionRBG}
\mathcal{S}_{RBG}&=&\frac{1}{2\kappa^2} \int d^4x \sqrt{-g} \mathcal{L}_G(g_{\mu\nu},R_{\mu\nu}(\Gamma)) \\
&+& \int d^4x \sqrt{-g} \mathcal{L}_m(g_{\mu\nu},\psi_m) \ , \nonumber
\end{eqnarray}
with the following definitions and conventions: $\kappa^2$ is Newton's constant in suitable units (in GR, $\kappa^2=8\pi G$), $g$ is the determinant of the space-time metric $g_{\mu\nu}$, which is \emph{a priori} independent on the affine connection $\Gamma \equiv \Gamma_{\mu\nu}^{\lambda}$ (metric-affine or Palatini formalism), from which the Ricci tensor follows as $R_{\mu\nu}(\Gamma) \equiv {R^\alpha}_{\mu\alpha\nu}(\Gamma)$. As we are dealing with minimally coupled bosonic fields in this work, torsion can be safely neglected \cite{Afonso:2017bxr}, and we can further keep just the symmetric piece of the Ricci tensor\footnote{These conditions guarantee the invariance of these theories under the so-called projective transformations, $\Gamma_{\mu\nu}^{\lambda} \to \Gamma_{\mu\nu}^{\lambda} + \xi_{\mu} \delta_{\nu}^{\lambda}$, where $\xi_{\mu}$ is an arbitrary one-form field associated to the gauge freedom in the parametrization of particle paths. This safeguards the theory against ghost-like degrees of freedom associated to the non-projectively invariant sector, see \cite{BeltranJimenez:2019acz,Jimenez:2020dpn} for details.} in the action (\ref{eq:actionRBG}). The RBG Lagrangian density $\mathcal{L}_G$ is built in terms of powers of traces of the object ${M^\mu}_{\nu} \equiv g^{\mu\alpha}R_{\alpha \nu}$ which includes, as particular examples, GR itself, $f(R)$ gravity, Ricci-squared gravities, EiBI gravity and its extensions \cite{BeltranJimenez:2017doy}, among many others. Finally, the matter Lagrangian $\mathcal{L}_m$ is assumed to depend on the metric and the matter fields $\psi_m$, but not on the connection, in order to ensure the  fulfillment of the equivalence principle \cite{WillsBook}.

It has been shown in \cite{BeltranJimenez:2017doy} that the field equations of any RBG defined by the action (\ref{eq:actionRBG}) can be written as
\begin{equation}\label{eq:GmnGeneral}
{G^\mu}_\nu(q)=\frac{\kappa^2}{|\hat{\Omega}|^{1/2}}\left[{T^\mu}_\nu-{\delta^\mu}_\nu\left(\mathcal{L}_G+\frac{T}{2}\right)\right] \ ,
\end{equation}
where $T_{\mu\nu} \equiv \frac{2}{\sqrt{-g}} \frac{\delta \mathcal{S}_m}{\delta g^{\mu\nu}}$ is the stress-energy tensor of the matter, and $T \equiv g^{\mu\nu}T_{\mu\nu}$ its trace. Here $q_{\mu\nu}$ represents the Einstein frame metric, which  is the one the independent connection is compatible with it, that is, $\nabla_{\mu}^{\Gamma} (\sqrt{-q}q^{\alpha\beta})=0$.  This metric is related to the space-time metric $g_{\mu\nu}$, which is the one the matter fields couple to (recall Eq.(\ref{eq:actionRBG})), via a relation of the form
\begin{equation}\label{eq:Omegafluid}
q_{\mu\nu}=g_{\mu\alpha}{\Omega^\alpha}_{\nu} \ ,
\end{equation}
where $\hat\Omega\equiv {\Omega^\alpha}_{\nu}$ is the \emph{deformation matrix} (hereafter a hat denotes a matrix and vertical bars its determinant), whose explicit form depends on the particular $\mathcal{L}_G$ chosen, but which can always be written \emph{on-shell} as a function of the matter fields and (possibly) the space-time metric as well (and the same applies to $\mathcal{L}_G$ itself).  The field equations (\ref{eq:GmnGeneral}) feature their second-order and gauge-free character, recover the GR solutions in vacuum (since in such a case ${T^\mu}_{\nu}=0$ one has ${R^\mu}_{\nu}(q)=0$ and $q_{\mu\nu}=g_{\mu\nu}$, modulo a trivial constant re-scaling), and propagate two tensorial polarizations of the gravitational field (gravitational waves) traveling at the speed of light.

\subsection{Mapping with anisotropic fluids}

The form of the field equations (\ref{eq:GmnGeneral}) compellingly invite us to rewrite them under the standard Einsteinian form of GR, that is
\begin{equation} \label{eq:GRframe}
{G^\mu}_\nu(q)=\kappa^2 \bar{T}{^\mu}_{\nu}(q) \ ,
\end{equation}
where from (\ref{eq:GmnGeneral}) the new stress-energy tensor is given by
\begin{equation} \label{eq:mapping}
\bar{T}{^\mu}_{\nu}(q)=\frac{1}{\vert \hat \Omega \vert^{1/2}}\left[{T^\mu}_{\nu}-{\delta^\mu}_{\nu}\left(\mathcal{L}_G+\frac{T}{2}\right)\right] \ .
\end{equation}
If one could find a $\bar{T}{^\mu}_{\nu}(q)$ such that its dependence on $g_{\mu\nu}$ could be completely eliminated in favor of $q_{\mu\nu}$ and the matter fields, then a correspondence between the original RBG and GR would be established. To illustrate how this is possible, and to work with relative generality, let us consider the coupling of a generic RBG to a matter source represented by an anisotropic fluid stress-energy tensor of the form
\begin{equation} \label{eq:Tmunufluid0}
{T^\mu}_\nu=(\rho + p_{\perp}) u^{\mu}u_{\nu}+p_{\perp} {\delta^\mu}_{\nu} + (p_r-p_{\perp}) \chi^{\mu}\chi_{\nu} \ ,
\end{equation}
where normalized timelike $g_{\mu\nu}u^{\mu}u^{\nu}=-1$ and spacelike $g_{\mu\nu}\chi^{\mu}\chi^{\nu}=+1$ vectors have been introduced, while $\rho$ is the fluid energy density, $p_r$ its pressure in the direction of $\chi^{\mu}$, and $p_{\perp}(r)$ its tangential pressure in the direction orthogonal to $\chi^{\mu}$. As we will see, electric fields (among many other examples) can be described as anisotropic fluids of the form given above. This will facilitate our analysis of rotating electrically charged solutions.

Assuming now the existence of another anisotropic fluid on the GR frame (\ref{eq:GRframe}), given also by Eq.(\ref{eq:Tmunufluid0}), but with new functions $\{\rho^q,p_r^q,p_{\perp}^q\}$, {\it i.e.},
\begin{equation} \label{eq:Tmunufluid1}
\bar{T}{^\mu}_{\nu}=(\rho^q + p_{\perp}^q) v^{\mu}v_{\nu}+p_{\perp}^q {\delta^\mu}_{\nu} + (p_r^q-p_{\perp}^q) \xi^{\mu}\xi_{\nu} \ ,
\end{equation}
for new timelike, $q_{\mu\nu}v^{\mu}v^{\nu}=-1$, and spacelike, $q_{\mu\nu}\xi^{\mu}\xi^{\nu}=+1$ vectors, then the relations  (\ref{eq:mapping}) explicitly formalize into
\begin{eqnarray}
p_{\perp}^q&=&\frac{1}{\vert \hat{\Omega} \vert^{1/2}} \left[\frac{\rho-p_r}{2} -\mathcal{L}_G \right]  \label{eq:mapflu1} \\
\rho^q+p_{\perp}^q&=&\frac{\rho+p_{\perp}}{\vert \hat{\Omega} \vert^{1/2}}   \label{eq:mapflu2} \\
p_r^q-p_{\perp}^q&=&\frac{p_r-p_{\perp}}{\vert \hat{\Omega} \vert^{1/2}}  \ . \label{eq:mapflu3}
\end{eqnarray}
These scalar relations must be supplemented with the relations $u^{\mu}u_{\nu}=v^{\mu}v_{\nu}$ and $\chi^{\mu}\chi_{\nu} =\xi^{\mu}\xi_{\nu}$. One can easily verify that for an anisotropic fluid and a given RBG Lagrangian, ${\Omega^\mu}_{\nu}$ in Eq.(\ref{eq:Omegafluid}) can always be written as
\begin{equation}  \label{eq:Omegafluid_EF}
{\Omega^\mu}_{\nu}=\alpha {\delta^\mu}_{\nu} + \beta u^{\mu}u_{\nu} + \gamma \chi^{\mu}\chi_{\nu} \ ,
\end{equation}
or, equivalently, as
\begin{equation}  \label{eq:Omegafluid_JF}
{\Omega^\mu}_{\nu}=\alpha {\delta^\mu}_{\nu} + \beta v^{\mu}v_{\nu} + \gamma \xi^{\mu}\xi_{\nu} \ ,
\end{equation}
where $\alpha$, $\beta$, and $\gamma$ are functions of the variables $\rho, p_r, p_{\perp}$ or, equivalently, of $\rho^q, p^q_r, p^q_{\perp}$. The explicit form of all these functions depends on the specific RBG Lagrangian that one considers.

\subsection{Example: EiBI gravity with electromagnetic fields}

To work out an explicit scenario, let us consider EiBI gravity, whose action can be expressed under the compact form \cite{BeltranJimenez:2017doy}
\begin{equation} \label{eq:actionEiBI}
\mathcal{S}_{EiBI}= \frac{1}{\kappa^2 \epsilon} \int d^4x \left(\sqrt{-q} - \lambda \sqrt{-g}\right) \ .
\end{equation}
Here $q_{\mu\nu} \equiv g_{\mu\nu} + \epsilon R_{\mu\nu}(\Gamma)$ denotes the connection-compatible metric in this case, $\nabla_{\mu}^{\Gamma}(\sqrt{-q}q^{\alpha\beta})=0$, and $\epsilon$ is a parameter with dimensions of length squared,  such that a perturbative expansion in series of $\vert R_{\mu\nu} \vert \ll 1/\epsilon $ brings (\ref{eq:actionEiBI}) into
\begin{eqnarray}
\mathcal{S}_{EiBI} &\approx& \int d^4x \sqrt{-g} \left(\frac{R}{2\kappa^2} - \Lambda_{eff}\right) \\
&-& \frac{\epsilon}{4\kappa^2} \int d^4x \sqrt{-g} \left( -\frac{R^2}{2}+R_{\mu\nu}R^{\mu\nu} \right) + \mathcal{O}(\epsilon^2)  \ , \nonumber
\end{eqnarray}
which is nothing but GR with an effective cosmological constant $\Lambda_{eff}=\frac{\lambda-1}{\kappa^2 \epsilon}$, supplemented by quadratic curvature corrections. For the action (\ref{eq:actionEiBI}) the deformation matrix is determined by the relation  \cite{BeltranJimenez:2017doy}
\begin{equation} \label{eq:OmegaEiBI}
\vert \hat{\Omega} \vert^{1/2} {(\Omega^{-1})^\mu}_{\nu}=\lambda {\delta^\mu}_{\nu} -\kappa^2 \epsilon {T^\mu}_{\nu} \ ,
\end{equation}
while the EiBI Lagrangian above can be conveniently expressed as
\begin{equation} \label{eq:LEiBIos}
\mathcal{L}_G=\frac{\vert \hat{\Omega} \vert^{1/2}-\lambda}{\kappa^2\epsilon} \ .
\end{equation}
Using Eq.(\ref{eq:OmegaEiBI}) and the expression for the stress-energy tensor (\ref{eq:Tmunufluid0}), one finds
\begin{eqnarray} \label{eq:OmegaEiBI_JF}
\vert \hat{\Omega} \vert^{1/2} ({\Omega^\mu}_{\nu})^{-1}&=&(\lambda-\epsilon\kappa^2p_{\perp}) {\delta^\mu}_{\nu}  \\ & - &\kappa^2 \epsilon \left[(\rho+p_{\perp})u^\mu u_\nu+(p_r-p_{\perp})\chi^\mu \chi_\nu\right] \ , \nonumber
\end{eqnarray}
which can also be written in terms of the Einstein frame variables using Eqs.(\ref{eq:mapflu1}), (\ref{eq:mapflu2}) and (\ref{eq:mapflu3})  as
\begin{eqnarray} \label{eq:OmegaEiBI_EF}
({\Omega^\mu}_{\nu})^{-1}&=&\left(1-\frac{\epsilon\kappa^2}{2}[\rho^q-p^q_r]\right) {\delta^\mu}_{\nu}  \\ & - &\kappa^2 \epsilon \left[(\rho^q+p_{\perp}^q)v^\mu v_\nu+(p^q_r-p^q_{\perp})\xi^\mu \xi_\nu\right]\ . \nonumber
\end{eqnarray}
Using the fundamental relation (\ref{eq:Omegafluid}), one finds the space-time metric in the RBG frame as
\begin{eqnarray} \label{eq:gqEiBI}
g_{\mu\nu}&=& \left(1-\frac{\epsilon\kappa^2}{2}[\rho^q-p^q_r]\right) q_{\mu\nu}  \\ & - &\kappa^2 \epsilon \left[(\rho^q+p_{\perp}^q)v_\mu v_\nu+(p^q_r-p^q_{\perp})\xi_\mu \xi_\nu\right]\ . \nonumber
\end{eqnarray}
This last relation is extremely powerful as it provides a solution for the EiBI theory starting from any known solution in GR supported by an anisotropic fluid source.

Since the matter Lagrangian densities generated on either RBG and GR sides will be typically non-linear, let us now busy ourselves with non-linear electrodynamics (NED). These can be defined in terms of a Lagrangian density $\psi(X,Y)$, where $X=-\frac{1}{2}F_{\mu\nu}F^{\mu\nu}$ and $Y=\frac{1}{2} F_{\mu\nu}F^{*\mu\nu}$ are the two field invariants that can be built out of the field strength tensor $F_{\mu\nu}=\partial_{\mu}A_{\nu}-\partial_{\nu}A_{\mu}$ and its dual $F^{*\mu\nu}=\frac{1}{2}\epsilon^{\mu\nu\alpha\beta}F_{\alpha\beta}$, where $A_{\mu}$ is the vector potential. Spherically symmetric electrostatic fields have a single non-vanishing component $F_{tr} \neq 0$ (thus $Y=0$) and can be read off as fluids satisfying $p_r=-\rho$ and $p_{\perp}=K(\rho)$, where the function $K(\rho)$ characterizes (implicitly) the corresponding electrodynamics theory via the identifications \cite{Afonso:2018mxn}
\begin{eqnarray}
\psi(X)&=&8\pi K(\rho) \label{eq:NED1} \\
\psi-2X \psi_X&=&-8\pi \rho \label{eq:NED2}  \ .
\end{eqnarray}
Now, the mapping equations (\ref{eq:mapflu1}), (\ref{eq:mapflu2}) and (\ref{eq:mapflu3}) for this combination of EiBI gravity (\ref{eq:actionEiBI}) with electrostatic fields establish that (here tildes denote an implicit factor $\epsilon\kappa^2$) \cite{Afonso:2018mxn}
\begin{eqnarray}
\tilde{\rho}_{\text{\tiny BI}}&=&\frac{\lambda\tilde{\rho}_{\text{\tiny GR}}-(\lambda-1)}{1-\tilde{\rho}_{\text{\tiny GR}}} \label{mapned1} \\
\tilde{K}_{\text{\tiny BI}}&=&\frac{\lambda \tilde{K}_{\text{\tiny GR}}+(\lambda-1)}{1+\tilde{K}_{\text{\tiny GR}}} \ . \label{mapned2}
\end{eqnarray}
These equations imply that, in general, any matter described by a NED on the RBG side, as defined by $\tilde{K}_{\text{\tiny BI}}$, will be mapped into a \emph{different} NED on the GR side, $\tilde{K}_{\text{\tiny GR}}$. This fact can be proved in general, irrespective of the restriction to spherical symmetry \cite{OORip}. To be more concrete, let us consider standard Maxwell electrodynamics on the GR side, $\psi_{\text{\tiny GR}}(X)=X$, which satisfies $\tilde{K}_{\text{\tiny GR}}=\tilde{\rho}_{\text{\tiny GR}}$. Thus, Eqs.(\ref{mapned1}) and (\ref{mapned2}) imply that (from now one we shall restrict ourselves to asymptotically flat solutions, $\lambda=1$)
\begin{equation}
\tilde{K}_{\text{\tiny BI}}=\frac{\tilde{\rho}_{\text{\tiny BI}}}{1+2\tilde{\rho}_{\text{\tiny BI}}} \ .
\end{equation}
In order to get the NED associated to this fluid, we write the relations (\ref{eq:NED1}) and (\ref{eq:NED2}) under the explicit form
\begin{eqnarray}
\psi &=& 8\pi \frac{{\rho}_{\text{\tiny BI}}}{1+2\tilde{\rho}_{\text{\tiny BI}}}  \\
\psi - 2X \psi_X &=&  - 8\pi \rho_{\text{\tiny BI}}  \ .
\end{eqnarray}
The solution to this system of equations is
\begin{equation}\label{eq:NED_EiBI}
{\psi(X)}=\frac{4\pi}{\kappa^2 \epsilon}\left(1-\sqrt{1- \frac{\kappa^2\epsilon}{2\pi} X}\right) \ ,
\end{equation}
where an integration constant has been set to get Maxwell electrodynamics in the $\epsilon \to 0$ limit, that is, $\underset{\epsilon \to 0}{\text{lim}} \psi(X) \approx X$. It is worth noting the resemblance between this NED and the well known Born-Infeld theory of electrodynamics, which is of the form \cite{BI34}
\begin{equation}
\psi_{\text{\tiny BI}}(X)=2\beta^2\left(1-\sqrt{1-\frac{X}{\beta^2}}\right) \ .
\end{equation}
By identifying $\beta^2\to 2\pi/\kappa^2\epsilon$ the correspondence is exact. One should note that the sign of $\epsilon$ is not necessarily positive, whereas $\beta^2$ is typically seen as positive in NED scenarios.

Let us further elaborate on the result above. In order to generate a rotating solution in the EiBI gravity theory starting with known solutions in GR, for the sake of this paper we wish to use the Kerr-Newman solution. That solution involves a rotating Maxwell electric field, which is equivalent to a rotating anisotropic fluid with $\tilde{K}_{\text{\tiny GR}}=\tilde{\rho}_{\text{\tiny GR}}$. Mapping this solution into the EiBI theory implies coupling it to the Born-Infeld-type NED of Eq.(\ref{eq:NED_EiBI})\footnote{Let us point out that, though for the sake of this paper we have explicitly shown this correspondence between these two theories for the case of spherically symmetric solutions, indeed, it can be proven \cite{Orazi:2020mhb} that it holds no matter the symmetry of the space-time background. This guarantees that we can safely use it when dealing with axially symmetric solutions in next section.}. To be precise, we find a correspondence between
\begin{equation} \label{eq:actionM1}
\mathcal{S}_{GR+Max}=\frac{1}{2\kappa^2}\int d^4x \sqrt{-g}R+\frac{1}{8\pi}\int d^4x \sqrt{-g}X \ ,
\end{equation}
and
\begin{eqnarray} \label{eq:actionM2}
\mathcal{S}_{EiBI+BI}&=&\frac{1}{\kappa^2 \epsilon }\int d^4x \left[\sqrt{-|g_{\mu\nu}+\epsilon R_{\mu\nu}|}- \sqrt{-g}\right]\nonumber \\ &+&\frac{1}{8\pi}\int d^4x \sqrt{-g}\psi_{BI}(X) \ .
\end{eqnarray}
In some previous works on electrostatic fields within EiBI gravity \cite{Olmo:2013gqa} we had assumed that $\epsilon<0$, which implies that the corresponding NED has the {\it wrong} sign as compared to the standard Born-Infeld NED.  In Ref.\cite{Olmo:2013mla} the case of quadratic gravity\footnote{In the context of NEDs, quadratic gravity (in the Ricci and Ricci-squared scalars) and EiBI gravity yield exactly the same electrostatic solutions due to the peculiar behaviour of the powers of the electromagnetic stress-energy tensor in that case, see \cite{Afonso:2018mxn} for a detailed discussion of this point.} coupled to Born-Infeld NED was studied but that NED has the opposite sign to that presented here. Let us point out that, though some aspects of the scenario considered here were analyzed in Ref. \cite{BI2}, our considerations in the next section will present complementary information to the one presented there and, moreover, it will allow to compare the results obtained by direct attack in that work and via the mapping introduced here.  This scenario will help us to build  confidence on the reliability of the mapping procedure before addressing the rotating case.

\section{Spherically symmetric solutions of EiBI gravity coupled to Born-Infeld NED} \label{sec:III}

\subsection{Generating the solutions via direct calculation}

The field equations for a free NED field take the form $\partial_\mu\left(\sqrt{-g}\psi_X F^{\mu\nu}\right)=0$ which, for static, spherically symmetric electric fields lead to $X \psi_X^2=Q^2/r^4$, where $Q$ is a constant identified as the electric charge. We are assuming a line element for $g_{\mu\nu}$ of the form
\begin{equation}\label{eq:gmn}
d{s}_g^2=-A(r)dt^2+{B(r)}dr^2+r^2(d\theta^2+\sin^2\theta d\varphi^2) \ .
\end{equation}
For the NED model (\ref{eq:NED_EiBI}),
one finds that the NED equation can be solved as
\begin{equation}\label{eq:XEiBI}
X=\frac{Q^2}{r^4+4s r_c^4}\ ,
\end{equation}
where we have introduced the scale $r_c^4\equiv  |\epsilon|\kappa^2 Q^2/8\pi$ and $s$ denotes the sign of $\epsilon$, that is $\epsilon=s \vert \epsilon \vert$.   On the other hand, the stress-energy tensor associated to a given NED takes the form
\begin{equation}
{T^\mu}_\nu=\frac{1}{8\pi}\left(\begin{array}{cc} (\psi-2X\psi_X)\hat{I}_{2\times 2} &  \hat{0}_{2\times 2}\\  \hat{0}_{2\times 2}& \psi \hat{I}_{2\times 2}\end{array}\right) \ ,
\end{equation}
where $\hat{I}$ and $\hat{0}$ are the $2 \times 2$ identity and zero matrices, respectively. Upon replacement of the Born-Infeld NED model (\ref{eq:NED_EiBI}) and the field   (\ref{eq:XEiBI}), the above expression becomes
\begin{equation}
{T^\mu}_\nu=\left(\begin{array}{cc} T_+\hat{I}_{2\times 2} & \hat{0}_{2\times 2}\\  \hat{0}_{2\times 2} & T_- \hat{I}_{2\times 2}\end{array}\right) \ ,
\end{equation}
where
\begin{eqnarray}
T_+&=&+\frac{s Q^2}{16\pi r_c^4} \left(1-\frac{\sqrt{r^4+4s r_c^4}}{r^2}\right) \\
T_-&=& -\frac{s Q^2}{16\pi r_c^4} \left(\frac{r^2}{\sqrt{r^4+4s r_c^4}}-1\right) \ .
\end{eqnarray}
It is easily seen that the above expressions recover the Maxwell ones when expanded in series of $r_c\to 0$. Now, replacing them into the definition of the deformation matrix (\ref{eq:OmegaEiBI}), it also inherits a structure of $2 \times 2$ blocks of the form
\begin{equation} \label{eq:Omegaem}
{\Omega^\mu}_\nu=\left(\begin{array}{cc} \Omega_+\hat{I}_{2\times 2} & \hat{0}_{2\times 2}\\  \hat{0}_{2\times 2} & \Omega_- \hat{I}_{2\times 2}\end{array}\right) \ ,
\end{equation}
where we have
\begin{eqnarray}\label{eq:Omegap}
\Omega_+&=& \frac{1}{2} \left(1+\frac{r^2}{\sqrt{r^4+4 s r_c^4}}\right) \\
\Omega_-&=& \frac{1}{2} \left(1+\frac{\sqrt{r^4+4 s r_c^4 }}{r^2}\right)\label{eq:Omegam} \ .
\end{eqnarray}
The RBG field equations (\ref{eq:GmnGeneral}) are thus written in this case as
\begin{equation}\label{eq:Rmn-generic}
{R^\mu}_\nu(q)=\frac{\kappa^2}{|\hat{\Omega}|^{1/2}}\left(\begin{array}{cc} (\LL_G+T_+) \hat I_{2\times2} & \hat 0 \\ \hat 0 & (\LL_G+T_-) \hat I_{2\times2} \end{array}\right) \ ,
\end{equation}
where $|\hat{\Omega}|^{1/2}=\Omega_+ \Omega_-$, with the above expressions for $T_{\pm}$ and $\Omega_{\pm}$, while for the EiBI Lagrangian $\mathcal{L}_G$ reads as in Eq.(\ref{eq:LEiBIos}). Since in the expression (\ref{eq:Rmn-generic}) the right-hand-side contains the radial coordinate of the metric  $g_{\mu\nu}$, while the left-hand side the one of the metric $q_{\mu\nu}$, we need next to work out the relation between them. To this end, we introduce the line element for the $q_{\mu\nu}$ metric as
\begin{equation}\label{eq:hmn-concrete1}
d{s}^2_q=-C(x)e^{2\psi(x)}dt^2+\frac{1}{C(x)}dx^2+x^2(d\theta^2+\sin^2\theta d\varphi^2) \ ,
\end{equation}
Comparing the angular sectors of (\ref{eq:hmn-concrete1}) and (\ref{eq:gmn}) via the fundamental relation (\ref{eq:Omegafluid}) with the structure (\ref{eq:Omegaem}) for this case, one finds $x^2=\Omega_- r^2$, which  leads to the explicit expressions
\begin{eqnarray}\label{eq:r(x)}
r^2&=&\frac{x^4-s r_c^4}{x^2} \\
x^2&=&\frac{1}{2}\left(r^2+\sqrt{r^4+4s r_c^4}\right) \ . \label{eq:x(r)}
\end{eqnarray}

On the other hand, from the structure in $2 \times 2$ blocks of the field equations (\ref{eq:Rmn-generic}) one can consider the substraction ${R^t}_t(q)-{R^x}_x(q)=0$ yielding $\psi(x)=$constant, which can be put to zero by a redefinition of the time coordinate without loss of generality. Introducing now the standard mass ansatz
\begin{equation}\label{eq:C(x)}
C(x)=1-\frac{2M(x)}{x} \ ,
\end{equation}
the ${R^\theta}_\theta$ component of the field equations (\ref{eq:Rmn-generic}) leads to
\begin{equation}\label{eq:Mx-general}
\frac{2M_x}{x^2}=\frac{\kappa^2}{\sqrt{|\Omega|}} (\LL_G+T_-) \ ,
\end{equation}
which,  in terms of the variable $r$, can be written, using (\ref{eq:r(x)}), as
\begin{eqnarray}\label{eq:Mr-general}
M_r&=&\frac{\kappa^2\Omega_-^{1/2}}{2\Omega_+} (\LL_G+T_-) r^2\left[1+\frac{r}{2}\frac{\Omega_{-,r}}{\Omega_-}\right] \\
&=&  \frac{\kappa ^2 Q^2}{8 \pi  \sqrt{2} }\frac{ r \sqrt{r^2+\sqrt{r^4+4 r_c^4 s}}}{\left(r^2 \left(r^2+\sqrt{r^4+4r_c^4 s}\right)+4r_c^4 s\right)} \ .\nonumber
\end{eqnarray}
The integration of this function is straightforward and leads to
\begin{equation}\label{eq:Mp}
M(r)=M_0 -\frac{\kappa ^2 Q^2}{8 \pi  \sqrt{2} }\frac{1}{\sqrt{r^2+\sqrt{r^4+s 4 r_c^4}}} \ ,
\end{equation}
where $M_0$ is an integration constant that can be identified with the asymptotic Schwarzschild mass. It is worth noting that using (\ref{eq:x(r)}) the above mass function turns into
\begin{equation}\label{eq:Mx}
M(r(x))=M_0 -\frac{\kappa ^2 Q^2}{16 \pi x}  \ ,
\end{equation}
which is nothing but the Reissner-Nordstr\"om mass function in terms of the radial coordinate $x$. Finally, the full expression for the space-time metric can be worked out using again (\ref{eq:Omegafluid}) as
\begin{equation}\label{eq:gmn-F}
d{s}^2=-\frac{C(x)}{\Omega_+}dt^2+\frac{dx^2}{C(x)\Omega_+}+r^2(x)(d\theta^2+\sin^2\theta d\varphi^2) \ ,
\end{equation}
where $C(x)$ is given by (\ref{eq:C(x)}) for the mass function (\ref{eq:Mx}) and $r^2(x)$ is given in (\ref{eq:r(x)}), while the $\Omega_\pm$ functions (\ref{eq:Omegap}) and (\ref{eq:Omegam}) can be directly expressed in terms of the coordinate $x$ as
\begin{eqnarray}\label{eq:Omegapx}
\Omega_+&=& \frac{1}{1+\frac{s r_c^4}{x^4}} \\
\Omega_-&=& \frac{1}{1-\frac{s r_c^4}{x^4}}  \ . \label{eq:Omegamx}
\end{eqnarray}
This completes our construction of the solutions by direct calculation. Note that setting $ r_c \rightarrow 0$, we recover the Reissner-Nordstr\"om solution of GR, as  expected.

\subsection{Generating the solutions via the mapping}\label{sec:RNviaMap}

We will now use the final mapping equation presented in (\ref{eq:gqEiBI}) to derive the above solutions  using the Reissner-Nordstr\"om solution of GR as the seed one. The Einstein-frame metric $q_{\mu\nu}$ is thus given by Eq.(\ref{eq:hmn-concrete1}) with the metric functions
\begin{equation} \label{eq:RNGR}
\psi=0 \hspace{0.2cm};\hspace{0.2cm} C(x) = 1-\frac{r_S}{x}+\frac{r_q^2}{2 x^2} \ ,
\end{equation}
where $r_S \equiv 2M_0$ is the Schwarzschild radius, and $r_q^2=\kappa^2 Q^2/(4\pi)$ the charge radius. For a Maxwell field, in the fluid view its components
are given by
\begin{eqnarray}
\rho^q&=&\frac{r_q^2}{2\kappa^2 x^4} \label{eq:rho_Maxwell}\\
p_{\perp}^q &=& \rho^q =-p_r^q \label{eq:prhop}\\
v_\mu&=& (-C(x)^{1/2},0,0,0) \\
\xi_\mu&=& (0,C(x)^{-1/2},0,0)  \ .
\end{eqnarray}
With these definitions and using Eq.(\ref{eq:OmegaEiBI_EF}), for a Maxwell field the $\Omega_\pm$ functions take the form
\begin{eqnarray}
\Omega_+&=& \frac{1}{1+\epsilon\kappa^2\rho^q} \\
\Omega_-&=& \frac{1}{1-\epsilon\kappa^2\rho^q}  \ .
\end{eqnarray}
One can readily check that these expressions coincide with those given in (\ref{eq:Omegapx}) and (\ref{eq:Omegamx}) when $\rho^q$ is written using (\ref{eq:rho_Maxwell}). It is thus immediate to verify that the metric functions in (\ref{eq:gmn}) read $A(r(x))=C(x)/\Omega_+$, $B(r(x))=(dx/dr)^2/(C(x)\Omega_+)$, and $r^2(x)=x^2/\Omega_-$, in complete agreement with the line element (\ref{eq:gmn-F}) and the definitions introduced so far.

This verifies that not only does Maxwell theory coupled to GR, as defined by action (\ref{eq:actionM1}), maps into a Born-Infeld-type NED coupled to EiBI gravity, as defined by action (\ref{eq:actionM2}), but also are their solutions mapped into each other. This has just been checked for the electrostatic configurations. Let us study next the properties of the configurations just obtained, from which valuable lessons will be learned before addressing the highly involved structure of their rotating counterparts.

\subsection{Properties of the solutions}

\begin{figure}[t!]
	\includegraphics[width=0.45\textwidth]{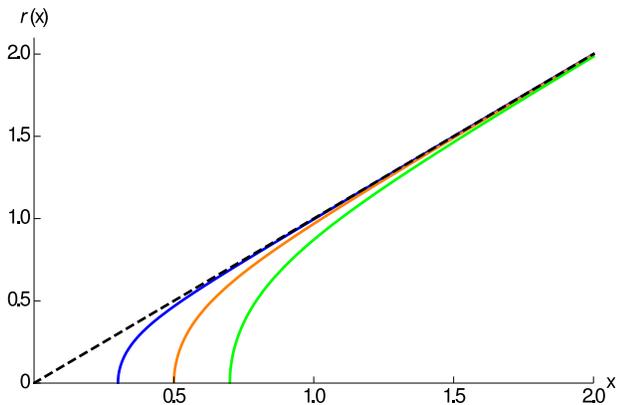}
	\caption{Radial function $r(x)$ for the case $ s=+1 $. From left to right, the curves represent $r_c=0$ (GR case, $r=x$, dashed black) $ r_c = 0.3$ (solid blue), $ r_c= 0.5 $ (solid orange) and $ r_c=0.7 $ (solid green). On each case, when $ x=r_c $, the radial function $ r(x) $ vanishes, and the solutions cannot be further extended beyond this point.}\label{r_pos}
\end{figure}

The relation of Eq.\eqref{eq:r(x)} between the radial functions in the $q_{\mu\nu}$ and $g_{\mu\nu}$ geometries depends on the sign of $ \epsilon $. By direct inspection of this equation for the case $ \epsilon>0$ (equivalently, $ s =+1$), $r(x)$ is positive only if $x>r_c$ as shown in Fig.\ref{r_pos}. Therefore, the area of the two-spheres, $A=4\pi r^2(x)$, is only positive in that region and, consequently, the geometry is only defined for $x\ge r_c$, with $x=r_c$ representing its center.

\begin{figure}[t]
\includegraphics[width=0.45\textwidth]{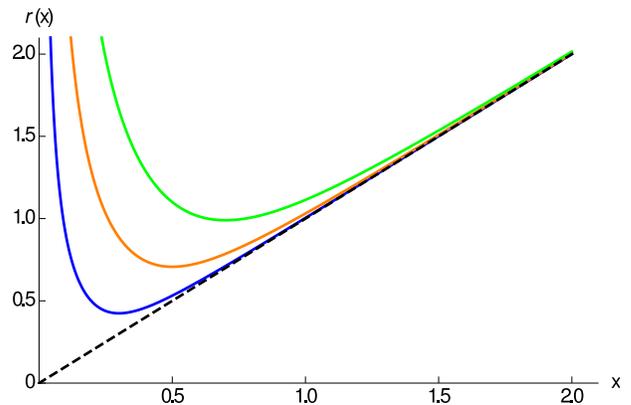}
\caption{Radial function $r(x)$ for the case $ s=-1 $. Same notation as in Fig.\ref{r_pos}. In this case a bounce at $x=r_c$ occurs for every value of $r_c$, signalling the presence of the throat of a wormhole.}\label{r_neg}
\end{figure}

For the case $ s=-1 $, $ r(x) $ has a minimum at $x=r_c$, growing again in the interval $0 \leq x<r_c$, and becoming infinite as $x\to 0$. The bounce in the $r(x)$ function signals the existence of a wormhole structure \cite{Visser}, with its throat located at $x=r_c$ (where $r=\sqrt{2}r_c$), as is apparent from Fig.\ref{r_neg}. Two comments are in order. First, as opposed to other wormholes found in other Palatini settings \cite{Olmo:2016tra}, here the coordinate $x$ does not extend over the entire real line, which means that we are dealing with a kind of asymmetric wormhole, which interpolates from asymptotic infinity to the central region of these objects. Second, given the fact that for scales much larger than $r_c$, it is easy to see that (\ref{eq:gmn-F}) recovers the Reissner-Nordstr\"om  solution (\ref{eq:RNGR}) of GR because $\Omega_\pm\to 1$, this means that  our solution will only depart from the well known GR results at those scales in which  $\Omega_{\pm}$
become important (this last statement is true for both $s=\pm 1$ branches). Therefore, astrophysical-size solutions of this kind would presumably look very much like the GR charged black hole but by tiny corrections, and with a wormhole  lurking on its interior.

\subsubsection{Horizons}

By direct calculation, one finds that the norm of vectors normal to the $x=$ constant hypersurfaces, $n^\mu\equiv g^{\mu\nu}\partial_\nu x$, is given by
\begin{equation}\label{eq:normal}
n^\mu n_\mu =\frac{x^2 \left(r_q^2/2+ x^2-r_S x\right)}{\left(x^4+s r_c^4\right)} \ ,
\end{equation}
which, in general, vanishes at
\begin{equation}
x_{\pm}^{RN}=\frac{r_S\pm \sqrt{r^2_S-2r_q^2}}{2}=M\pm \sqrt{M^2-\frac{\kappa^2Q^2}{8\pi}} \label{hor_res},
\end{equation}
where $x=x_{\pm}^{RN}$ represents the location of the usual horizons of the Reissner-Nordstr\"om solution. The metric component $ g_{tt} (x) $ defined in Eq.\eqref{eq:gmn-F} can be explicitly expressed in terms of the coordinate $x$ as
\begin{equation}\label{eq:gttsph}
g_{tt}(x)=-\left(1+s\frac{r_c^4}{x^4}\right)\left(1-\frac{r_S}{x}+\frac{r_q^2}{2x^2}\right) \ ,
\end{equation}

\begin{figure}[t]
	\includegraphics[width=0.45\textwidth]{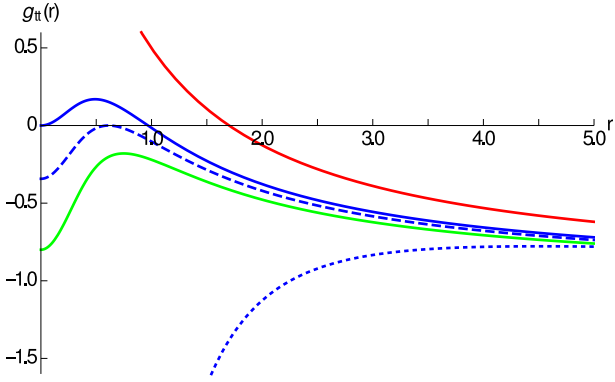}
	\caption{Metric component $ g_{tt} (r)$, for the case $ s=+1 $ and $r_c$ = 0.5. One finds i) Reissner-Nordstr\"om-like solutions with two (blue solid, $r_q$ = 1, $r_s$ = 1.5), a single degenerated (blue dashed, $r_q$ = 1, $r_s$ = $ \sqrt{2} $) and no horizons (blue dotted, $r_q$ = 3, $r_s$ = 2); ii) Schwarzschild-like solutions with a single horizon (red, $r_q$  = 1, $r_s$  = 2);  iii) solutions with no horizons (green, $r_q$ = 0.5, $r_s$ = 1.3). All solutions are asymptotically flat, $g_{tt} \rightarrow -1$ as $r \rightarrow \infty$.}
	\label{gtt_pos}
\end{figure}

Let us consider first the case $ \epsilon>0 $ ($s=+1$). Since the term inside the first parenthesis in (\ref{eq:gttsph})  is always finite and positive, the structure of horizons is controlled by the (Reissner-Nordstr\"om-like) term inside the second parenthesis, where we bear in mind that since $x \geq r_c$ this term will be always finite as well and, moreover, its roots will be given by $x_{\pm}^{RN}$ in Eq.(\ref{hor_res}). Therefore, there are several possible configurations depending on the choice of the black hole parameters and the EiBI/BI parameter $ \vert \epsilon \vert$. The discussion simplifies if one writes the location of the horizons in terms of $r(x)$ as $r_\pm=\sqrt{(x_{\pm}^{RN})^4-r_c^4}/x_{\pm}^{RN}$. This clearly shows that when $ x^{RN}_{\pm} > r_c $, one finds the presence of Reissner-Nordstr\"om type solutions with two horizons, a single but degenerate horizon (corresponding to extreme black holes) or no horizons (naked solutions). For $ x^{RN}_{-} < r_c < x^{RN}_{+}$, the solutions are Schwarzschild-like black holes, which have a single non-degenerate horizon. Finally, for $ x^{RN}_{\pm} < r_c  $ they have no horizons. For all of them the metric component $g_{tt}(r)$ is finite at $ r=0 $ for every $r_c \neq 0$, as depicted in Fig.\ref{gtt_pos}. It is worth mentioning that this structure of horizon mimics the one of certain families of NED models coupled to GR \cite{DiazAlonso:2009ak}.

\begin{figure}[t!]
	\includegraphics[width=0.46\textwidth]{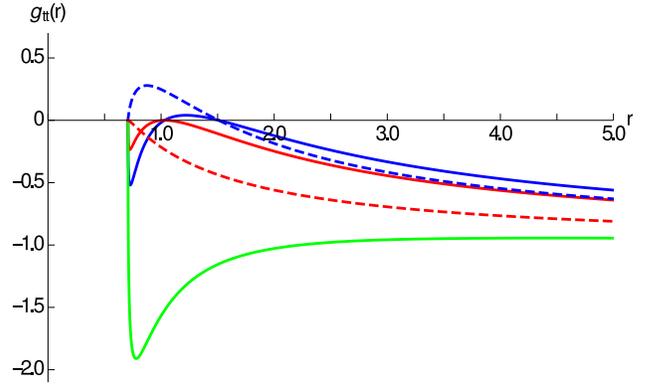}
	\caption{Metric component $ g_{tt} (r) $, for the case $ s=-1 $ and $ r_c = 0.5 $ and values of  $ r_q=\sqrt{3} ,r_s = 2.5 $ (solid blue); $ r_q=\sqrt{1.5}, r_s= 2 $ (dashed blue);  $ r_q=\sqrt{2}, r_s = 2 $ (solid red); $ r_q=1/\sqrt{2} , r_s = 1 $ (dashed red) and $ r_q=3/2, r_s = 1/2 $ (green). }
	\label{gtt_neg}
\end{figure}

For the case $ s=-1 $, we first note that $ g_{tt} (x) $ vanishes always at $ x=r_c $, as it follows directly from Eq.(\ref{eq:gttsph}). However, it is unclear if such a point is  a null hypersurface or not because rather than vanishing, the norm (\ref{eq:normal}) there is divergent.  Given that the time-like Killing vector $\partial_t$ always has vanishing norm at $x=r_c$ and that $n^\mu n_\mu$ changes sign across this hypersurface, it is advisable to pay more careful attention to our choice of coordinates. In order to do so, we rewrite Eq.(\ref{eq:gmn-F}) in terms of Eddington-Finkelstein coordinates as
\begin{eqnarray}\label{eq:gmn-F_EF}
d{s}^2&=&-\left(1+\frac{sr_c^4}{x^4}\right)C(x)dv^2+2\left(1+\frac{sr_c^4}{x^4}\right)dvdx\nonumber \\&+&r^2(x)(d\theta^2+\sin^2\theta d\varphi^2) \ .
\end{eqnarray}
It is now evident that for $s=-1$ the metric is singular at $x=r_c$, which might be the cause of the ill-defined norm of the $x=$constant hypersurfaces precisely there. We can thus assume that $x$ is not a good coordinate at $x=r_c$ so that a new coordinate $y(x)$ satisfying $\pm\left(1-\frac{r_c^4}{x^4}\right)dx=dy$ could fix the problem in the metric, with the plus sign corresponding to $x>r_c$ and the minus to $x<r_c$ to guarantee the monotonicity of $y$. The corresponding change of coordinates is then given by 
\begin{equation} \label{eq:y(x)EF}
 y=\left\{\begin{array}{lr} x+\frac{r_c^4}{3x^3}\quad \quad \quad \quad \quad  \text{ if } x\ge r_c \\
 \frac{8}{3}r_c-\left(x+\frac{r_c^4}{3x^3}\right)  \quad  \text{ if } 0<x\le r_c \end{array}\right. \ ,
 \end{equation}
and turns (\ref{eq:gmn-F_EF}) into
\begin{eqnarray}\label{eq:gmn-F_EF2}
d{s}^2&=&-\left(1-\frac{r_c^4}{x(y)^4}\right)C[x(y)]dv^2+2dvdy\nonumber \\&+&\left(x(y)^2+\frac{r_c^4}{x(y)^2}\right)(d\theta^2+\sin^2\theta d\varphi^2) \ ,
\end{eqnarray}
which now is nonsingular at $y(x=r_c)=4r_c/3$. One can check that the norm of the vector normal to the $y=\text{constant}$ hypersurfaces\footnote{It should be noted that this is equivalent, up to a constant, to that of the hypersurfaces of constant area $A=4\pi x^2(1+r_c^4/x^4)$.} is now given by 
\begin{equation}\label{eq:normalsm}
\tilde{n}^\mu \tilde{n}_\mu =\left(1-\frac{r_c^4}{x(y)^4}\right)\frac{ (x(y)-x_+^{RN})(x(y)-x_-^{RN})}{2x^2} \ ,
\end{equation}
and clearly vanishes at $x=x_\pm^{RN}$ and at $x=r_c$, with the latter being unavoidable as long as there is charge.  As a comparison with the $s=+1$ case, in Fig.\ref{gtt_neg}  we have plotted the metric $g_{tt} $ as a function of the radial function $r(x)$, where we observe that this metric component is zero at the wormhole throat, $x=r_c$ or $r=\sqrt{2}r_c$, as expected.

\subsubsection{Curvatures}\label{sec:curvatures}

 A glance at the behavior of curvature scalars provides additional useful information. In fact, for $\epsilon>0$ one can check that the Ricci and Ricci-squared scalars diverge at $x=r_c$, the former as $R\sim  1/(x-r_c)^2$ and the latter as $R_{\mu\nu}R^{\mu\nu}\sim 1/(x-r_c)^4$. For $\epsilon<0$, one finds instead $R\sim  1/(x-r_c)^3$ and  $R_{\mu\nu}R^{\mu\nu}\sim 1/(x-r_c)^6$.

Interestingly, that generic behavior can be softened by a specific choice of parameters. Indeed, if we set
\begin{equation} \label{eq:rSa}
r_S\equiv r_c+\frac{r_q^2}{2r_c} \ ,
\end{equation}
then for both signs of $\epsilon$ we get $R\sim  1/(x-r_c)$ and $R_{\mu\nu}R^{\mu\nu}\sim 1/(x-r_c)^2$. Moreover, for that choice of $r_S$, one finds that
\begin{equation}
x_-^{RN}=r_c \quad \text{and} \quad x_+^{RN}=\frac{r_q^2}{2r_c}
\end{equation}
with $r_S=r_c+x_+^{RN}>r_c$. Note that these quantities are completely determined by the charge (recall that $r_q^2\equiv \kappa^2 Q^2/4\pi$ and $r_c^4\equiv |\epsilon| r_q^2/2$). When $s=+1$, then we have that $x_+^{RN}$ represents an outer horizon, while $x=x_-^{RN}$ represents an inner horizon located right at the center of the object (vanishing area). For $s=-1$, a glance at Eq.(\ref{eq:normalsm}) shows that the normal vector at $x=x_-^{RN}=r_c$ has a  vanishing norm and, therefore, it can be regarded as an inner horizon. These solutions are thus characterized by two (event and inner) horizons. Given that $r_c^4\equiv |\epsilon| r_q^2/2$, we have  that for this family of solutions $x_-^{RN}\sim r_q^{1/2}$ and $x_+^{RN}\sim r_q^{3/2}$ grow with the charge at different rates, which may lead to configurations with $x_-^{RN}>x_+^{RN}$. Thus, the wormhole throat may lie inside the horizon, if $x_+^{RN}>r_c$, or outside, if $x_+^{RN}<r_c$. The equality occurs when the extremality condition $x_+^{RN}=x_-^{RN}=r_c$ is satisfied, which implies $r_c^2=r_{q}^2/2$ or, equivalently, $r_{q}\equiv r_{q}^{ext}=\sqrt{2|\epsilon|}$.

Turning back to the curvatures, one finds that taking the additional constraint
\begin{equation} \label{eq:spcialcase}
r_q=\sqrt{\frac{22}{17}}r_c=\frac{11}{17} r_{q}^{ext} \ ,
\end{equation}
in the $\epsilon<0$ ($s=-1$) case yields completely regular curvature invariants (including the Kretschmann scalar), while no choice of $r_q$ can cure the Ricci and Ricci-squared (or Kretschmann) curvature scalars simultaneously in the $\epsilon>0$ ($s=+1$) case. It should be noted that this regular configuration of the $s=-1$ case is to be regarded as microscopic, since $|\epsilon|$ is expected to be very small and we are assuming that (\ref{eq:rSa}) can be at least of order $r_c$.

As a final comment, we note that the curvature scalars as $x\to 0$ in the $\epsilon<0$ case are also finite  there, taking the values $R=36/|\epsilon|$, $R_{\mu\nu}R^{\mu\nu}=396/\epsilon^2$, and ${R^\alpha}_{\beta\mu\nu} {R_\alpha}^{\beta\mu\nu}= 408/\epsilon^2$, respectively.

\subsubsection{Surface gravity}

From a thermodynamic point of view, it is remarkable that  the surface gravity of the usual horizons takes the form
\begin{equation}\label{eq:kappa}
\kappa_\pm=\frac{x_\pm^{RN}-x_\mp^{RN}}{2(x_\pm^{RN})^2} \ ,
\end{equation}
regardless of the sign and value of $\epsilon$. This means that the temperature of these horizons, $T_\pm=\kappa_\pm/4\pi$, is exactly the same as in GR.  For those configurations satisfying Eq.(\ref{eq:rSa}), the discussion depends on the sign of $\epsilon$ as follows. 

\begin{figure}[t!]
	\includegraphics[width=0.45\textwidth]{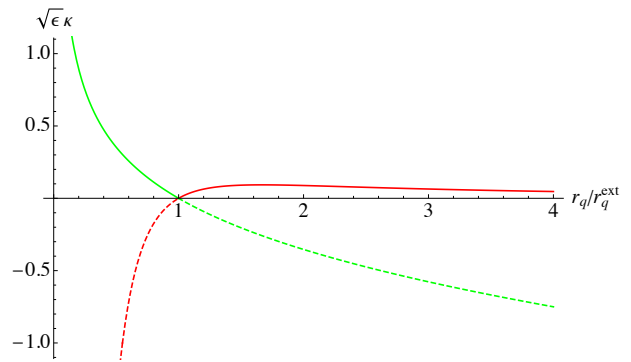}
	\caption{Representation of the surface gravity when Eq.(\ref{eq:rSa}) is fulfilled. The (green) curve represents the inner horizon temperature, $\kappa_-$, while the red one is the outer horizon, $\kappa_+$. Note that $\kappa_+=\kappa_-=0$ in the extremal case ($r_q=\sqrt{2 \vert \epsilon \vert}$). Recall that for this choice of $r_S$ the inner horizon is fixed at the location $x=r_c$. }.\label{Fig:Tp-Tm}
\end{figure}

If $s=+1$, then $x_-^{RN}=r_c$ is a null hypersurface (of vanishing area, see Eq.(\ref{eq:r(x)})) with an associated temperature given by (\ref{eq:kappa}). The outer/inner horizon has positive/negative temperature until $r_q=\sqrt{2|\epsilon|}\equiv r_q^{ext}$, which defines the zero temperature degenerate configuration. For smaller values of $r_q$, the outer horizon disappears and it is the inner one (with vanishing area and fixed at $x_-^{RN}=r_c$) which develops a positive temperature. This temperature grows unbounded as the charge goes to zero and $r_c$ shrinks, which indicates that the extremal configuration must be an equilibrium point [see  Fig.\ref{Fig:Tp-Tm}].  One can verify that  the entropy associated  to this horizon stays constant at the value $S=\pi r_c^2$ regardless of the temperature, which diverges as $\sim 1/\sqrt{r_q}$.  Obviously, there is no natural statistical interpretation of this entropy in terms of microscopic states.

When $s=-1$, we saw that $x=r_c$ also represents an event horizon. In this case, the surface gravity takes the general form
\begin{equation}
\kappa_c=\lim_{x\to r_c}\frac{2r_c^2+r_q^2-2r_cr_S}{4r_c^2(x-r_c)} \ ,
\end{equation}
which is always divergent except when the condition (\ref{eq:rSa}) is satisfied. In that case, it leads to $\kappa_c=(r_c-r_q^2/2r_c)/r_c^2$, which can also be expressed as $\kappa_c=(x_-^{RN}-x_+^{RN})/r_c^2$, recovering in this way the general formula (\ref{eq:kappa}). This result further reinforces the exceptional character of the configurations satisfying the constraint (\ref{eq:rSa}). Note that the zero temperature case occurs when $x_+^{RN}$  coincides with the wormhole throat.

\subsubsection{Geodesics}

To gain further insight on the geometry of these solutions, let us now focus on their geodesic structure. The geodesic equation for a static spherically symmetric space-time \cite{Olmo:2016tra} can be particularized to the present case as follows:
\begin{equation}
\left(1+\frac{s r_c^4}{x^4} \right)^2 \left(\frac{dx}{du}\right)^2=E^2-V_{eff} \ ,
\end{equation}
where $u$ is the affine parameter and $V_{eff}$ the effective potential, which is explicitly obtained as
\begin{equation} \label{eq:Veff}
V_{eff}=\left(1+\frac{s r_c^4}{x^4}\right)\left(1-\frac{r_S}{x} +\frac{r_q^2}{2x^2} \right) \left(\frac{L^2x^2}{x^4-s r_c^4}-k\right) \ ,
\end{equation}
where $E$ and $L$ are the energy per unit mass and angular momentum per unit mass, respectively, and $k=0,-1$ for null and time-like geodesics, respectively. For null ($k=0$) radial ($L=0$) geodesics this equation can be exactly integrated as
\begin{equation} \label{eq:nullradial}
x \left(1-\frac{s r_c^4}{3x^4}\right)=\pm E(u-u_0) \ ,
\end{equation}
where the $\pm$ sign corresponds to outgoing/ingoing geodesics, and $u_0$ is an integration constant. For large distances, $x \to \infty$, one finds the GR result, $ x = \pm E(u-u_0)$.

\begin{figure}[t!]
	\includegraphics[width=0.45\textwidth]{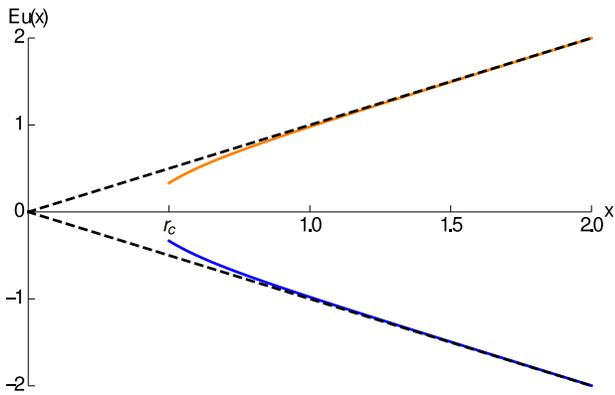}
	\caption{The affine parameter $E \cdot u$ versus the radial coordinate $x$ for ingoing (blue) and outgoing (orange) null radial geodesics in the case $s=+1$ with $ r_c = 0.5 $ and $ u_0 =0 $.  These geodesics reach the surface $x=r_c$ in finite affine time and end there, since there is no further space-time these geodesics could be continued to. The black dashed lines correspond to null radial geodesics for GR, which are also incomplete.}
	\label{fig:nullgeopos}
\end{figure}

In the case $s=+1$, the surface $x=r_c$ is reached in finite affine time, like $x=0$ in the GR case and, therefore, these solutions are said to be geodesically incomplete because they hit a curvature divergence and there is no possible further extension beyond this point [see Fig.\ref{fig:nullgeopos}].

\begin{figure}[t!]
	\includegraphics[width=0.45\textwidth]{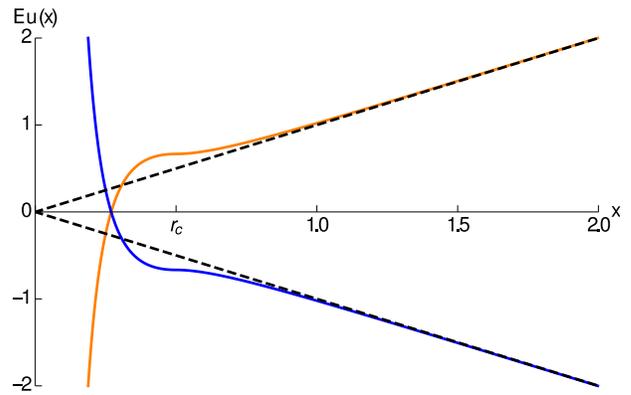}
	\caption{The affine parameter $E \cdot u$ versus the radial coordinate $x$ for ingoing (blue) and outgoing (orange) radial null geodesics in the case $s=-1$, setting $ r_c= 0.5 $ and $u_0=0$. Since the range of $u(x)$ covers the whole real axis, these geodesics are complete. The GR result is also plotted (black dashed), to compare it with the results of our geometry.}\label{Fig:nullgeo}
\end{figure}

When $s=-1$, the situation requires a bit more of attention in order to make sure that outgoing geodesics have $dx/du\ge0$ and ingoing geodesics  $dx/du\le 0$ all over their domains. In this regard, for outgoing geodesics we have
\begin{eqnarray}
\left( \frac{r_c^4}{x^4}-1\right)\frac{dx}{du}&=&E \ \ \text{ if } x\le r_c \\
\left(1- \frac{r_c^4}{x^4}\right)\frac{dx}{du}&=&E \ \ \text{ if } x\ge r_c  \ ,
\end{eqnarray}
whose integration leads to
\begin{equation}
E \cdot u(x)=\left\{ \begin{array}{lr}
\frac{8r_c}{3}-x\left(1+\frac{r_c^4}{3x^4}\right) & \text{ if } x\le r_c \\
x\left(1+\frac{r_c^4}{3x^4}\right) & \text{ if } x\ge r_c
\end{array}\right.
\end{equation}
For ingoing geodesics we only have to change the sign in front of $E$ above. As can be seen from the above solution, as $x\to \infty$ we have $u(x)\to +\infty$, while as $x\to 0$ we get $u(x)\to -\infty$, which confirms that these geodesics would be complete [see Fig.\ref{Fig:nullgeo}] if we were allowed to match them at the wormhole throat, were curvature scalars generically diverge. In this respect, it is important to note that these radial null geodesics are insensitive to the dependence of the metric on $r_q$ and $r_S$.  Note also that, as we saw above in Sec. \ref{sec:curvatures}, there is a specific choice of parameters for which no curvature divergences arise at the throat, see Eq.(\ref{eq:spcialcase}). Since there is no reason to suspect that null radial geodesics are incomplete in that case, and the geodesic equation is the same for all cases, it seems legitimate to accept their completeness in all the $s=-1$ cases. As a final remark in this regard, one should note that even in the case in which the Ricci, Ricci-squared, and Kretschman scalars are finite, it is always possible to build new {\it ad hoc} scalars such that divergences may arise at any point that we may wish. Given that there is not an obvious or privileged category of fundamental scalars, it is unclear what the implications of the divergence of particular curvature scalars for the extendibility of geodesics might be.

The discussion of timelike and nonradial null geodesics presents a richer phenomenology that will be presented elsewhere.

\section{A rotating black hole solution} \label{sec:IV}

\subsection{Derivation of the solution}

Let us now busy ourselves with the derivation of the rotating black hole solution, which is the main result of this work. Attempting to obtain rotating solutions via direct resolution of the field equations is a highly non-trivial challenge for any modified theory of gravity, which one can only face once a reasonable ansatz has been found by some means, or via some suitable algorithm, like the Janis-Newman one \cite{Shaikh:2019fpu}. In our case we will use the same approach of the static case to map the charged rotating (Kerr-Newman) solution of GR into the corresponding one in the EiBI model, suitably adapting it to the particular structure of axisymmetric solutions. The process to achieve this goal goes as follows:

\begin{enumerate}
	\item Take the Kerr-Newman solution of GR in some coordinate system and compute the corresponding Einstein tensor.
	\item Interpret that solution as generated by the stress-energy tensor of an anisotropic fluid. A direct comparison of the Kerr-Newman Einstein tensor with the stress-energy tensor of the fluid provides concrete expressions for the fluid density/pressure functions and the fluid's angular velocity $\omega$.
	\item Use those expressions to generate the corresponding solution in the EiBI theory using the relations (\ref{eq:gqEiBI}).
\end{enumerate}

Before proceeding we first note that, unlike in the Reissner-Nordstr\"om case, where the vector $v^\mu$ only has the temporal component, for the Kerr-Newman solution the fluid velocity vector has a spatial component, namely (assuming motion in the $\phi$ direction only)
\begin{equation}
v^\mu=(v^t,0,0,v^\phi) \ ,
\end{equation}
where typically $v^\phi$ is written as $v^\phi=\omega v^t$, with $\omega$ representing the angular velocity, \emph{i.e.}, $\omega \equiv d\phi/dt$. The normalization of the vector $v^\mu v^\nu q_{\mu\nu}=-1$ implies that
\begin{equation}
v^t=\frac{1}{\sqrt{-(q_{tt}+\omega q_{t\phi}+\omega^2 q_{\theta \phi})}} \ ,
\end{equation}
but it does not constraint the form of $\omega$, which must be set by other means. In this case, $\omega$ represents the angular velocity of observers comoving with a fluid that mimics the same external field as a rotating electric charge. Recall that for a Maxwell electric field, the density and pressures of the equivalent fluid satisfy Eq.(\ref{eq:prhop}). The normalization of the spatial vector $\xi^\mu$ is enough to fully constrain its only spatial component, $\xi^\mu=(0,\xi^x,0,0)$, giving $\xi^x=1/\sqrt{q_{xx}}$. Thus, from the comparison of the Einstein tensor and the fluid stress-energy tensor we will need to solve for the energy density $\rho$ and the angular velocity $\omega$.

Let us now write the Kerr-Newman solution of GR using ($t,x,\theta,\phi$)  coordinates in Boyer-Lindquist form as \cite{WaldBook}
\begin{equation}\label{eq:KNinBL}
q_{\mu\nu}= \left(
\begin{array}{cccc}
-(1-f) & 0 & 0 & -a f \sin ^2\theta \\
0 & \frac{\Sigma }{\Delta } & 0 & 0 \\
0 & 0 & \Sigma  & 0 \\
-a f \sin ^2\theta & 0 & 0 &  \left(x^2+a^2+f a^2\sin ^2\theta\right)\sin ^2\theta \\
\end{array}
\right)
\end{equation}
where $0 \leq a \leq 1$ is the spin parameter and the following definitions apply
\begin{eqnarray}
f&=& \frac{r_S x-r_q^2/2}{\Sigma}=\frac{x^2+ a^2-\Delta}{\Sigma} \nonumber \\
\Sigma&=& x^2+ a^2\cos^2\theta \label{eq:defsBL}\\
\Delta&=& x^2-r_S x+a^2+r_q^2/2  \nonumber \ .
\end{eqnarray}
The corresponding Einstein tensor for this line element can be conveniently written  as
\begin{equation}
{G^\mu}_\nu= \left(
\begin{array}{cccc}
\frac{-r_q^2\left(x^2 +a^2[1 +  \sin^2\theta]\right)}{2\Sigma^3} & 0 & 0 & \frac{a r_q^2 \left(a^2+r^2\right) \sin ^2\theta }{\Sigma^3} \\
0 & \frac{-r_q^2}{2\Sigma^2} & 0 & 0 \\
0 & 0 & \frac{r_q^2}{2\Sigma^2} & 0 \\
\frac{ -a r_q^2}{\Sigma^3} & 0 & 0 & \frac{r_q^2 \left(x^2 +a^2[1+ \sin^2\theta]\right)}{2\Sigma^3} \\
\end{array}
\right)
\end{equation}
and, according to Eq.(\ref{eq:Tmunufluid1}), the fluid  stress-energy tensor becomes
\begin{equation}
{T^\mu}_\nu= \left(
\begin{array}{cccc}
-\frac{\rho^q}{F_1} & 0 & 0 & F_2 \\
0 & -\rho^q & 0 & 0 \\
0 & 0 & \rho^q & 0 \\
-F_3 & 0 & 0 & \frac{\rho^q}{F_1}  \\
\end{array}
\right) \ ,
\end{equation}
where $F_1$, $F_2$, and $F_3$ are complicated expressions of the fluid and metric variables, becoming unity in the non-rotating limit, $a \to 0, \omega  \to  0$. For instance, the function $F_1$ takes the form
\begin{eqnarray}
F_1&=&\frac{\sin ^2\theta \left(\omega ^2 \left(a^2+x^2\right)^2+a^2-2 a f \Sigma  \omega \right)}{\Delta  \left(1-a^2 \omega ^2 \sin ^4\theta\right)+\sin ^2\theta \left(\omega ^2 \left(a^2+x^2\right)^2-a^2\right)} \nonumber \\
&-&\frac{\Delta  \left(a^2 \omega ^2 \sin ^4\theta+1\right)}{\Delta  \left(1-a^2 \omega ^2 \sin ^4\theta\right)+\sin ^2\theta \left(\omega ^2 \left(a^2+x^2\right)^2-a^2\right)} \ , \nonumber
\end{eqnarray}
By direct comparison between the expressions of ${G^\mu}_\nu$ and $\kappa^2{T^\mu}_\nu$ above, it is easy to see that
\begin{equation}\label{eq:densityKN}
\kappa^2\rho^q= \frac{r_q^2}{2\Sigma^2} \ ,
\end{equation}
which can be interpreted as the energy density of an axially symmetric electromagnetic field described by
\begin{equation}
A_{\mu}=(A_t,0,0,A_{\phi})=\frac{Qx}{\Sigma}(1,0,0,-a \sin^2 \theta) \ ,
\end{equation}
from which the components of the field strength tensor are immediately obtained in the usual way. Moreover, a glance at the above expression for $F_1$ shows that $\omega$ follows simply by solving a quadratic algebraic equation, whose physical solution (rotating in the same direction as the black hole) can be shown to correspond to
\begin{equation}
\omega= \frac{a}{a^2+x^2} \ .
\end{equation}

We thus now have all the necessary ingredients to map the Kerr-Newman solution of GR into a rotating solution in the EiBI gravity theory coupled to the Born-Infeld-type electrodynamics, as given by the action (\ref{eq:actionM2}). To this end, we parallel our approach to the static case presented in Section \ref{sec:RNviaMap} and apply it to the rotating case. For this purpose, we will use the metric (\ref{eq:KNinBL}) as the seed solution, together with the definitions (\ref{eq:defsBL}), and the following expressions
\begin{eqnarray}
\rho^q&=&\frac{r_q^2}{2\kappa^2 \Sigma^2} \ , \label{eq:rhoMaxwell}\\
p^q_{\perp} &=& \rho^q=-p^q_r \ , \\
v_\mu&=& \left(-\sqrt{\frac{\Delta}{\Sigma}},0,0,a \sin^2\theta \sqrt{\frac{\Delta}{\Sigma}}\right) \ , \\
\xi_\mu&=& \left(0,\sqrt{\frac{\Sigma}{\Delta}},0,0 \right)  \ ,
\end{eqnarray}
for the energy density $\rho^q$ and the radial $p_r^q$ and tangential $p_{\perp}^q$ pressures, and the unit vectors $v_{\mu}$ and $\xi_{\mu}$. Now, using the relation between metrics presented in (\ref{eq:gqEiBI}), the spacetime metric $g_{\mu\nu}$ can be written in this case as
\begin{equation}
g_{\mu\nu}=q_{\mu\nu}+\epsilon \kappa^2\rho^q h_{\mu\nu} \ ,
\end{equation}
where the additional piece induced by EiBI gravity corrections reads
	\begin{equation}
	h_{\mu\nu}=\left(\begin{array}{llll} -\frac{(\Delta+ a^2 \sin ^2\theta)}{\Sigma}& 0 & 0 &  \frac{a (\Delta+x^2+a^2)}{\Sigma} \sin ^2\theta \\
	0 &\frac{\Sigma}{\Delta} & 0 & 0 \\
	0 & 0 & -\Sigma & 0 \\
	\frac{a (\Delta+x^2+a^2)}{\Sigma} \sin ^2\theta & 0 & 0 & -\left[\frac{(x^2+a^2)^2+a^2\Delta \sin^2\theta}{\Sigma}\right]\sin ^2\theta
	\end{array}\right) \ .
	\end{equation}
Therefore, the final solution for a rotating black hole of EiBI gravity coupled to Born-Infeld-type electrodynamics (that is, the action of (\ref{eq:actionM2})) is obtained, in Boyer-Lindquist coordinates, as
\begin{widetext}
	\begin{eqnarray} \label{eq:axialline}
	ds^2&=&-\left(1-f +\epsilon \kappa^2 \rho^q \frac{(\Delta+ a^2 \sin ^2\theta)}{\Sigma}\right) dt^2
	- 2 a \left(f-\epsilon \kappa^2 \rho^q\frac{ (\Delta+x^2+a^2)}{\Sigma}\right) \sin ^2\theta dt d\phi  \\
	&+& \frac{(1+\epsilon \kappa^2 \rho^q)\Sigma}{\Delta} dx^2 +(1-\epsilon \kappa^2 \rho^q) \Sigma d\theta^2
	+  \Big[ \left(x^2+a^2+f a^2\sin ^2\theta\right)
	-\epsilon \kappa^2 \rho^q \frac{(x^2+a^2)^2+a^2\Delta \sin^2\theta}{\Sigma} \Big]   \sin ^2\theta d\phi^2 \ , \nonumber
	\end{eqnarray}
\end{widetext}
where the corrections in $\epsilon$ from EiBI gravity to the Kerr-Newman solution of GR are apparent. Pretty and clean. This is the main result of this work. 

A basic aspect to note is the fact that the vacuum solution, $r_q=0$ (which implies $\rho^q=p_r^q=p_{\perp}^q=0$), coincides with the Kerr solution of GR. This is consistent with the general statements about the behaviour of RBGs in absence of matter; only when the density function $\rho^q$ in Eq.(\ref{eq:rhoMaxwell}) is non-vanishing do we have modifications as compared to the GR solution. Indeed, far from the sources, the line element (\ref{eq:axialline}) boils down to
\begin{eqnarray}
ds^2&=&-\left[1-\frac{2M}{r}+\mathcal{O}(r^{-2})\right]dt^2 \\
&+&\left[\frac{4aM\sin^2 \theta}{r} +\mathcal{O}(r^{-2})\right]dtd\phi \nonumber \\
&+& \left[1+\mathcal{O}(r^{-1}) \right](dr^2+r^2d\Omega^2) \ ,
\end{eqnarray}
which is nothing but the asymptotic limit of an axially symmetric rotating body in Boyer-Lindquist coordinates. This implies the consistence of these objects with the observations of orbital motions of test particles around any of them. The changes induced by the RBG theory on GR solutions will be stronger in those regions where the energy density reaches its highest values, which in the present case is the innermost region (for the sake of this work, we shall leave aside the analysis of circular orbits outside of the event horizon, which deserves a separate analysis, to be carried out elsewhere).

Let us point out that, for slowly-rotating black holes, $a \ll 1$, the line element (\ref{eq:axialline}) boils down to
\begin{widetext}
	\begin{eqnarray} \label{eq:axiallineslowly}
	ds^2&=&\left[-\frac{(x^4+sr_c^4)\Delta_0}{x^6} + \Big(\frac{\cos^2 \theta \Delta_0}{x^4} + sr_c^4 \Big(-\frac{1+\sin^2 \theta}{x^6} + \frac{3\cos^2 \theta \Delta_0}{x^8} \Big) \Big) a^2 + \mathcal{O}(a^4) \right]dt^2 \nonumber \\
	&+& \left[  2a \sin^2 \theta  \left( \frac{r_q^2(x^4+sr_c^4) -2x(-2sr_c^4 x +r_S(x^4+sr_c^4))}{2x^6} \right) + \mathcal{O}(a^3) \right]dt d\phi \nonumber \\
	&+& \left[ \frac{x^4+sr_c^4}{x^2 \Delta_0} + \frac{-x^2(x^4+sr_c^4)+\cos^2 \theta (x^4-sr_c^4)  \Delta_0 }{x^2 \Delta_0^2}a^2 + \mathcal{O}(a^4)\right]dx^2
	\\
	&+& \left[ \left(x^2-\frac{sr_c^4}{x^4} \right) + \cos^2 \theta \left(1+\frac{sr_c^4}{x^4} \right) a^2 + \mathcal{O}(a^4) \right] d\theta^2 \nonumber  \\
	&+& \left[ \left(x^2-\frac{sr_c^4}{x^2} \right) + \frac{x^2(sr_c^4(-2+3\cos^2 \theta) +x^4) - \sin^2 \theta (r_q^2 (x^4+sr_c^4)/2 + x(sr_c^4x-r_S(x^4+sr_c^4))}{x^6}a^2 + \mathcal{O}(a^4) \right]d\phi^2 \ , \nonumber
	\end{eqnarray}
\end{widetext}
where $\Delta_0\equiv \Delta(a=0)=x^2-r_Sx+r_q^2/2$ and we have used that $\epsilon \kappa^2\rho^q\equiv s r_c^4/\Sigma^2$ via Eq.(\ref{eq:rhoMaxwell}). From this expression it is clearly seen that rotation induces linear and quadratic corrections in $a$ to the spherically symmetric solutions described in Sec.\ref{sec:III} and given by the line element (\ref{eq:gmn}).

Before concluding this section, for completeness we write the line element (\ref{eq:axialline}) in Eddington-Finkelstein coordinates given by the transformations $dv=dt+\frac{(x^2+a^2)}{\Delta}dx$ and $d\varphi=d\phi+a dx/\Delta$ as:
\begin{widetext}
	\begin{eqnarray} \label{eq:axialline-EF}
	ds^2&=&-\left(1-f +\epsilon \kappa^2 \rho^q \frac{(\Delta+ a^2 \sin ^2\theta)}{\Sigma}\right) dv^2 +2\left(1+\epsilon \kappa^2 \rho^q\right)dv dx
	- 2 a \sin ^2\theta\left(f-\epsilon \kappa^2 \rho^q\frac{ (\Delta+x^2+a^2)}{\Sigma}\right)  dv d\varphi  \\
	&-&2 a \sin ^2\theta\left(1+\epsilon \kappa^2 \rho^q\right)dx d\varphi +(1-\epsilon \kappa^2 \rho^q) \Sigma d\theta^2
	+  \Big[ \left(x^2+a^2+f a^2\sin ^2\theta\right)
	-\epsilon \kappa^2 \rho^q \frac{(x^2+a^2)^2+a^2\Delta \sin^2\theta}{\Sigma} \Big]   \sin ^2\theta d\varphi^2 \ \nonumber \ .
	\end{eqnarray}
\end{widetext}

\subsection{Properties of the solutions}

\subsubsection{Horizons and ergoregions}

In the rotating solution obtained above, one can verify that the vectors normal to the $x=$constant hypersurfaces have norm given by
\begin{equation}
n^\mu n_\mu=\frac{\Sigma(x-x_+^{KN})(x-x_+^{KN})}{\Sigma^2+sr_c^4} \ ,
\end{equation}
where we have defined
\begin{equation}
x_{\pm}^{KN}=\frac{r_S \pm \sqrt{r_S^2-4(a^2+r_q^2/2)}}{2} \ .
\end{equation}
From this expression, it is evident that the hypersurfaces $x=x_{\pm}^{KN}$ have vanishing norm, whereas there is a divergence if the denominator vanishes when $s=-1$. One can verify that this divergence is of the same nature as that found in the non-rotating case, since the hypersurface $(\Sigma-s r_c^4/\Sigma) =$constant has vanishing norm precisely at $\Sigma=r_c^2$ when $s=-1$.  The two hypersurfaces $x=x_{\pm}^{KN}$ turn out to be Killing horizons of the combination
\begin{equation}
\xi=\chi+\frac{a}{x^2+a^2}m \ ,
\end{equation}
where $\chi=\partial_t$ and $m=\partial_\phi$ are the Killing vectors associated to time translations and rotations around the symmetry axis, respectively. A glance at the norm of this vector $\xi$ puts forward that $\Sigma=r_c^2$ is also a Killing horizon when $s=-1$:
\begin{equation}
\xi^\mu \xi_\mu=\frac{(x-x_+^{KN})(x-x_+^{KN})(\Sigma^2+s r_c^4)}{(x^2+a^2)\Sigma^2} \ .
\end{equation}
Let us now consider the conditions for the existence of ergoregions. The norm of the  time-like Killing vector  $\chi$ is given by 
\begin{equation}\label{eq:norm_chi}
\chi^\mu \chi_\mu=\frac{(x-x^e_-)(x-x^e_+)}{(x^2+a^2\cos\theta^2)}+sr_c^4\frac{(x-y^e_-)(x-y^e_+)}{(x^2+a^2\cos\theta^2)^3} \ ,
\end{equation}
and it vanishes when the condition
\begin{equation}\label{eq:norm_ergo}
\frac{(x-x^e_-)(x-x^e_+)}{(x^2+a^2\cos\theta^2)}=-sr_c^4\frac{(x-y^e_-)(x-y^e_+)}{(x^2+a^2\cos\theta^2)^3} \ ,
\end{equation}
is satisfied, where we have introduced the definitions
\begin{eqnarray}
x^e_{\pm}&=&\frac{r_S \pm \sqrt{r_S^2-2r_q^2-4a^2\cos^2\theta}}{2} \\
y^e_{\pm}&=&\frac{r_S \pm \sqrt{r_S^2-2r_q^2-4a^2(\sin^2\theta+1)}}{2} \ .
\end{eqnarray}
In general, the solutions to Eq.(\ref{eq:norm_ergo}) are difficult to obtain analytically, though some useful information can be obtained on simple grounds. Firstly, note that
\begin{equation}
x^e_-\le y^e_-\le y^e_+\le x^e_+ \ ,
\end{equation}
and that the sign of the left-hand side of (\ref{eq:norm_ergo}) must be the same as that on the right-hand side. According to this, if $s=0$ (the GR case) then the only solutions are $x=x^e_\pm$, which define the usual ergoregions of the Kerr-Newman solution. If $s=+1$ then the solutions must satisfy
\begin{equation}
x^e_-\le x\le y^e_- \quad \text{and} \quad y^e_+\le x\le x^e_+ \ ,
\end{equation}
while if $s=-1$ then the solutions satisfy instead
\begin{equation}
x\le x^e_- \quad y^e_-\le x\le y^e_+ \quad \text{and} \quad x\ge x^e_+ \ .
\end{equation}
The equalities in the above relations are useful in order to note that, on the rotation axis ($\theta=0,\pi$), then $x^e_\pm=y^e_\pm$, which indicates that the new ergoregions are smooth deformations of those present in the GR solutions. Away from the rotation axis, a perturbative approach can be used to estimate the leading order corrections in some cases of interest. Given that in the $r_c\to 0$ limit one recovers the GR solutions $x^e_\pm$, approximated solutions can be obtained in the astrophysical limit, in which $r_c$ is much smaller than $x^e_-$ (assumed to be real). In fact, taking $x=x^e_\pm+\delta_\pm$ and using Eq.(\ref{eq:norm_ergo})  it is easy to verify that
\begin{equation}
\delta_\pm=\mp\frac{s r_c^4}{((x^e_\pm)^2+a^2\cos\theta^2)^2}\frac{(x^e_\pm-y^e_-)(x^e_\pm-y^e_+)}{(x^e_+-x^e_-)}+\mathcal{O}(r_c^8)
\end{equation}
In the slowly-rotating limit, where the line element is given by Eq.(\ref{eq:axiallineslowly}), this quantity boils down to
\begin{eqnarray}
\delta_\pm &\approx & \mp s\frac{ 8 a^2\sin\theta^2  r_c^4}{\sqrt{r_S^2-2r_q^2}\left(r_q^2+r_S\left(-r_S+\sqrt{r_S^2-2r_q^2}\right)\right)^2}\nonumber \\
&+&\mathcal{O}(a^4) \ .
\end{eqnarray}
This indicates that rotating black holes in extensions of GR can indeed exhibit an external ergoregion different from that expected in GR. Bounds on potential deviations could be obtained by the observation of accretion disks \cite{Bambi:2015kza}, shadows \cite{Wang:2018prk}, and other means \cite{Carballo-Rubio:2018jzw}, to be explored in future works. In this sense, we point out that by turning back to the exact expression (\ref{eq:norm_chi}), one can verify that on the hypersurface $\Sigma=r_c^2$ it boils down to $\chi^\mu \chi_\mu=2a^2\sin^2\theta/r_c^2$, which vanishes on the equatorial plane $\theta=\pi/2$. This provides an exact, though partial, solution to the existence of ergoregions, whose characterization will be carried out elsewhere. 

\subsubsection{Metric and curvature divergences}

We have already seen that the hypersurface $\Sigma=r_c^2$ leads to metric singularities when $s=-1$, as it induces the vanishing of the cross-term $dv dx$ in the representation (\ref{eq:axialline-EF}).  Additional problems can be guessed by looking at the zeros of the $g_{\theta\theta}$ component. From Eq.(\ref{eq:axialline}) there are, in principle, two such cases:
\begin{equation} \label{eq:ring}
\Sigma\equiv x^2+a^2\cos^2\theta=0 \hspace{0.2cm}; \hspace{0.2cm} \Sigma^2-sr_c^4=0 \ ,
\end{equation}
The first one corresponds to the region where the electromagnetic field energy density in GR diverges, see (\ref{eq:densityKN}), that is
\begin{equation}\label{eq:firstzero}
x=0 \hspace{0.2cm}; \hspace{0.2cm} \theta=\pi/2  \ .
\end{equation}
As for the second, it only admits a real solution in the branch $s=+1$, where it becomes
\begin{equation}\label{eq:secondzero}
\Sigma=r_c^2 \ \leftrightarrow x^2+a^2\cos^2\theta=r_c^2 \ .
\end{equation}
In the spherically symmetric limit $a\to 0$ studied in Sec. \ref{sec:III}, the $s=+1$ case that we are considering here was characterized by the vanishing of the area of the two-spheres at $x=r_c$, preventing $x$ from taking smaller values (since, otherwise, the metric signature would change). Now we see that a non-vanishing angular momentum has an important impact on this relation. As is evident, on the equatorial plane ($\theta=\pi/2$) this condition occurs exactly at $x=r_c$, like in the spherical case. Considering smaller values of $x$ on the plane $\theta=\pi/2$ would imply a change of signature in the metric and suggests that we should discard the first case in (\ref{eq:ring}), $\Sigma=0$, as unphysical for this value of $s$.  On the contrary, for $s=-1$, there is no signature change and $\Sigma=0$  still represents a physically acceptable region (a non-spherical boundary of infinite area).

Further useful information on the relevance of the conditions (\ref{eq:ring}) can be extracted by looking at curvature invariants. The Ricci scalar, for instance, has a structure of the form
\begin{equation}
R=\frac{P_s(x,\theta; a,rc^4,r_S,r_q^2)}{\Sigma \left(\Sigma-r_c^2 \right)^3} \ ,
\end{equation}
where $P_s(x,\theta; a,r_c^4,r_S,r_q^2)$ is a polynomial that vanishes if $r_c\to 0$. The denominator of this quantity shows the critical cases identified in (\ref{eq:ring}) and puts forward that the Ricci scalar has problems at $\Sigma=r_c^2$ regardless of the sign of $\epsilon$.  Something similar happens to $R_{\mu\nu}R^{\mu\nu}$ and   ${R^\alpha}_{\beta\mu\nu}{R_\alpha}^{\beta\mu\nu}$ but they have a much more complicated rational structure that does not illuminate the discussion.  We thus see that on the hypersurface $\Sigma^2=r_c^4$ curvature invariants diverge. This happens, in particular, when $x=r_c$ on the equatorial plane $\theta=\pi/2$, like in the spherically symmetric case.
As one moves back from $\theta=\pi/2$ towards $\theta\to 0$ (the rotation axis), the condition $\Sigma=r_c^2$ extends the range of $x$ below the spherical limit $x=r_c$. If $a^2>r_c^2$, we may reach $x=0$ at a critical angle $\theta_c$ such that
\begin{equation}
a^2\cos^2\theta_c=r_c^2 \ .
\end{equation}
For angles within $0\leq \theta < \theta_c$ and $\pi/2+\theta_c\leq \theta <\pi$, the condition $\Sigma=r_c^2$ cannot be satisfied.  In the limit $r_c^2/a^2 \ll 1$, it is clear that the divergence will be located very near the equatorial plane, being completely confined on the plane in the GR limit $r_c\to 0$ (yielding the well-known ring singularity).  If $a^2\leq r_c^2$ then the condition can be satisfied by some $0<x<r_c$ for any azimuthal angle $0\leq \theta \leq \pi$. A deeper analysis in terms (possibly of some extension) of Cartesian Kerr-Schild coordinates could help better understand the geometry of this singular region. However, together with its maximal analytical extension, this is a non-trivial aspect to be explored elsewhere because the general structure of the metric shown in (\ref{eq:gqEiBI})  indicates that $g_{\mu\nu}$ cannot be written in standard Kerr-Schild form even if $q_{\mu\nu}$ admits such a decomposition. The conformal factor in front of $q_{\mu\nu}$ is just one of the reasons against that possibility.

Before concluding this section, it should be noted that the energy density at $\Sigma=r_c^2$  is finite in the wormhole case ($s=-1$) but divergent in the other case ($s=+1$), as it follows from (\ref{mapned1}), because $\tilde{\rho}_{GR}\equiv \epsilon \kappa^2\rho^q=s r_c^4/\Sigma^2=s$ when $\Sigma=r_c^2$. The behavior of the angular pressures is just the opposite, as shown by  (\ref{mapned2}), being divergent in the wormhole case and finite in the other.  On the other hand, a glance at the induced geometry on the surfaces of $t$ and $x$ constant, shows that at $\Sigma=r_c^2$ the equatorial and polar circumferences (with $\theta=\pi/2$ and $\phi=$ constant, respectively) in the wormhole case  have proper length given by
\begin{equation}
l_{eq}=2\pi \sqrt{2}\frac{(r_c^2+a^2)}{r_c} \quad \text{and} \quad  l_{pol}=2\sqrt{2}\pi r_c \ ,
\end{equation}
which illustrates the lack of spherical symmetry due to the non-vanishing angular momentum.

\section{Conclusion} \label{sec:V}

Testing the Kerr hypothesis on the nature of astrophysical black holes via several means is currently a hot topic in the investigation on the reliability of General Relativity as compared to its many alternatives on their strong-field regime, thanks to the existence of a varied pool of present and future observations. However, the low supply of exact rotating solutions available in the literature of modified theories of gravity might prevent to test in detail the predictions of the modifications of GR using the full power of the observational machinery already available (see e.g. \cite{Carballo-Rubio:2018jzw} for a recent discussion on this topic). Therefore, the investigation of rotating black hole solutions of modified theories of gravity beyond GR demands the development of new strategies and ideas to address the specific challenges of this field.

In this paper we have put to work one such idea, namely, a recently discovered correspondence that allows to map the field equations of Ricci-based theories of gravity  formulated in metric-affine spaces and GR itself.  Via this correspondence any RBG coupled to some matter source can be mapped into GR coupled to the same matter source but described by a different Lagrangian density (non-canonical, in general), in such a way that the solutions of the former can be obtained from the solutions of the latter using purely algebraic transformations. This correspondence has been developed for general anisotropic fluids as a matter source though extensions to other scenarios (for instance, dynamical ones) are also possible. While in previous works the proof of concept of this method was established by re-deriving previously known spherically symmetric solutions, in the present work we have made used of this correspondence to show that GR coupled to Maxwell electrodynamics can be mapped into Eddington-inspired Born-Infeld gravity coupled to a Born-Infeld-type electrodynamics.

Equipped with this result we proceed to solve  the field equations of this setting for electrostatic, spherically symmetric scenarios by hard force, and then checked that the application of the mapping procedure yields exactly the same result by a much more direct and simpler route. Next we discussed the features of the configurations obtained this way for both signs of the EiBI parameter $\epsilon$. For the $s=+1$ case such configurations have a minimum radial coordinate $x=r_c$ beyond which no further extension of the space-time is possible, while in the $s=-1$ case this minimum radius represents the throat of a wormhole structure. These two different behaviors translate in a modified description of horizons, curvatures, and geodesics. Indeed, in the wormhole case, while curvature scalars generically diverge at the throat, the existence of a configuration with bounded scalars suggests that geodesics might be complete. 

Elaborating upon this firm ground, the main highlight of the present paper was to use this framework in order to find the rotating counterpart of the Kerr-Newman solution of GR in the EiBI theory. Doing this required a bit more care, but finally we managed to obtain this solution under a closed exact form (in both Boyer-Lindquist and Eddington-Finkelstein coordinates), where the new gravitational corrections to the GR solution appear as new terms in $\epsilon$. This solution has a number of distinctive properties, such as a different structure of horizons and ergoregions as compared to that of the Kerr-Newman one, a nontrivial surface with curvature divergences, a non-spherical (exact) wormhole structure, and other aspects related to the analytical extensions of the geometry and choices of coordinates  to be further explored elsewhere.

The bottom line of this discussion is the reliability of the mapping method to find exact analytical solutions of interest for metric-affine gravity of the kind considered here. The implementation of such solutions and the analysis of their features is very timely given the open opportunities available for testing modified gravity in astrophysical settings using the pool of available data: accretion disks, strong gravitational lensing and shadows, generation of gravitational waves in binary mergers, and so on. In this sense, despite all the limitations and its unlikely physical relevance, the exact rotating solutions analyzed here are very useful from the point of view of enriching the theoretical discussions regarding potentially observable deviations in the structure of such solutions as compared to GR ones. Now, exploring further families of nonlinear electrodynamics and/or scalar fields paralleling our analysis, and extensions of these results to other RBGs, could be useful to buid up a catalogue of new rotating solutions in more physically appealing models, in order to identify generic features and templates that can be compared with the data from  multimessenger astronomy. Work along these lines is currently underway.

\section*{Acknowledgments}
MG is funded by a predoctoral contract from the Comunidad de Madrid (Spain). GJO is funded by the Ramon y Cajal contract RYC-2013-13019 (Spain). DRG is funded by the \emph{Atracci\'on de Talento Investigador} programme of the Comunidad de Madrid (Spain) No. 2018-T1/TIC-10431, and acknowledges support from the Funda\c{c}\~ao para a Ci\^encia e a Tecnologia (FCT, Portugal) research grants Nos.  PTDC/FIS-OUT/29048/2017 and PTDC/FIS-PAR/31938/2017. This work is supported by the Spanish projects FIS2017-84440-C2-1-P (MINECO/FEDER, EU), i-LINK1215 (CSIC), SEJI/2017/042 (Generalitat Valenciana), the project H2020-MSCA-RISE-2017 Grant FunFiCO-777740,  the Severo Ochoa grant SEV-2014-0398 (Spain) and the Edital 006/2018 PRONEX (FAPESQ-PB/CNPQ, Brazil, Grant 0015/2019). This article is based upon work from COST Actions CA15117 and CA18108, supported by COST (European Cooperation in Science and Technology).  The authors thank A. Cardenas-Avendano for useful discussions and comments.

\end{document}